\documentclass[aps,pre,twocolumn,groupedaddress]{revtex4-2}

\usepackage{bm}
\usepackage{graphicx,color}
\usepackage[colorlinks=true,allcolors=blue]{hyperref}%
\hypersetup{
	pdftitle={Partition sum of thermal, under-constrained systems},
	pdfauthor={Cheng-Tai Lee and Matthias Merkel},
	pdfcreator={Cheng-Tai Lee and Matthias Merkel},
}
\usepackage{amsfonts}
\usepackage{amsmath}
\usepackage{accents}
\usepackage{mathtools}

\renewcommand{\d}{\mathrm{d}}
\newcommand{\vecn}[1]{\vec{#1}}
\newcommand{\vecd}[1]{\bm{#1}}
\newcommand{\underdot}[1]{{\underaccent{\dot}{#1}}}

\newcommand{\aref}[1]{\hyperref[#1]{appendix~\ref*{#1}}}

\makeatletter
\newcommand{\vast}{\bBigg@{3}}
\newcommand{\Vast}{\bBigg@{4}}
\makeatother

\definecolor{mmcolor}{rgb}{0.8, 0.0, 0.2}

\definecolor{ctcolor}{rgb}{0.8, 0.0, 0.2}

\allowdisplaybreaks

\begin{document}
\title{Partition sum of thermal, under-constrained systems}
\author{Cheng-Tai Lee}
\affiliation{Aix Marseille Univ, Universit\'e de Toulon, CNRS, CPT (UMR 7332), Turing Center for Living Systems, Marseille, France}
\author{Matthias Merkel}
\email[]{matthias.merkel@cnrs.fr}
\affiliation{Aix Marseille Univ, Universit\'e de Toulon, CNRS, CPT (UMR 7332), Turing Center for Living Systems, Marseille, France}


\begin{abstract}
Athermal (i.e.\ zero-temperature) under-constrained systems are typically floppy, but they can be rigidified by the application of external strain.
Following our recently developed analytical theory for the athermal limit, here and in the companion paper, we extend this theory to under-constrained systems at finite temperatures.
Close to the athermal transition point, we derive from first principles the partition sum for a broad class of under-constrained systems, from which we obtain analytic expressions for elastic material properties such as isotropic tension $t$ and shear modulus $G$ in terms of isotropic strain $\varepsilon$, shear strain $\gamma$, and temperature $T$.
These expressions contain only three parameters, entropic rigidity $\kappa_S$, energetic rigidity $\kappa_E$, and a parameter $b_\varepsilon$ describing the interaction between isotropic and shear strain. We provide analytical expressions for these parameters based on the microscopic structure of the system.
Our work unifies the physics of systems as diverse as polymer fibers \& networks, membranes, and vertex models for biological tissues.
\end{abstract}

\maketitle

\section{Introduction}
Under-constrained systems possess more degrees of freedom (dofs), $N_\mathrm{dof}$, than constraints (or springs), $N_s$.
In the athermal limit, i.e.\ at zero temperature, under-constrained systems are generally floppy. Conversely, over-constrained systems, which have $N_\mathrm{dof}<N_s$, are generally rigid \cite{Maxwell1864,Calladine1978}. 
We recently developed a generic analytic framework to predict the elastic properties of athermal, under-constrained systems \cite{Merkel2019,Lee2022}. However, such a framework does not yet exist for finite temperatures.

In the athermal limit, under-constrained systems are typically floppy due to the existence of zero modes.  For instance, there are \emph{infinitesimal} zero modes, i.e.\ collective displacements of the dofs that do not affect any of the constraints \emph{to linear order}. The number of linearly independent infinitesimal zero modes, $N_0$, is given by \cite{Maxwell1864,Calladine1978}:
\begin{equation}
	N_0 = N_\mathrm{dof}-N_s+N_{SSS},\label{eq:N0}
\end{equation}
where $N_{SSS}$ is the number of linearly independent states of self-stress (SSS).
A SSS is a combination of virtual tensions put on the springs of a network that does not result in any net force on the dofs.
According to Eq.~\eqref{eq:N0}, under-constrained systems always have a positive number of infinitesimal zero modes.

There are three possibilities regarding the nature of infinitesimal zero modes.
First, they could be \emph{trivial} zero modes, i.e.\ global translations or rotations of the system, which leave the ``shape'' of the system unchanged.
Second, they could correspond to \emph{mechanisms}, i.e.\ \emph{finite} displacements that change the ``shape'' of the system, but do not affect any of the constraints \cite{Lubensky2015,Mannattil2022}.
Third, infinitesimal zero modes -- while inducing no changes to the constraints to linear order -- could induce changes to higher order, corresponding to higher-order rigidity \cite{Damavandi2022,Damavandi2022a}. Here, we refer to this third kind as \emph{$K$th-order zero modes} with some integer $K>0$. Such modes do not affect the constraints to $K$th order, but do so to $(K+1)$th order (precise definitions in \autoref{sec:model - zero modes}).
Because many of the infinitesimal zero modes often correspond to mechanisms, at zero temperature, under-constrained systems are typically floppy.

Still, under-constrained systems can be rigidified by externally applied strain, such as isotropic expansion or shear \cite{Calladine1978,Alexander1998,Cui2019b,Merkel2019,Damavandi2022,Damavandi2022a,Zhang2022a}.
This is because at some amount of externally applied strain, the constraints cannot be fulfilled any more. For a spring network, this happens when the amount of external strain becomes geometrically incompatible with the spring rest lengths. 
Right at the rigidity transition, a SSS is created \cite{Lubensky2015}. This allows prestresses to form once the network is strained any further.
Beyond the transition point both SSS and prestresses rigidify the system \cite{Calladine1978,Alexander1998,Wyart2005b,Lubensky2015,Merkel2018,Merkel2019}.
In previous work, we used these insights to analytically predict the elastic material properties close to the rigidity transition in the athermal limit \cite{Merkel2019,Lee2022}.
These ideas apply not only to spring networks, but also to other systems such as polymer networks \cite{Onck2005,Broedersz2014,Licup2015,Sharma2016a} and vertex models for biological tissues \cite{Moshe2017,Merkel2018,Sussman2018,Merkel2019,Wang2020}.

Despite these advances for the athermal limit, there is still relatively little known about \emph{thermal}, i.e.\ finite-temperature, under-constrained systems \cite{Plischke1998, Dennison2013, Wigbers2015}. Recent work used effective-medium theory (EMT) to obtain analytical expressions for the shear modulus of spring networks \cite{Zhang2016a,Chen2023}, and Mannattil et al.\ discuss the effect of singularities in the configuration space on the mechanics of the system \cite{Mannattil2022}.
However, we still lack a generic theory for the elastic properties of thermal, under-constrained systems.

Here, we analytically derive from first principles the partition sum and elastic properties of thermal, under-constrained systems close to the athermal transition for small isotropic strain $\varepsilon$, shear strain $\gamma$, and temperature $T$.
After defining the class of under-constrained systems that we discuss here (\autoref{eq:model}), we first revisit the athermal limit (\autoref{sec:athermal limit}). This will provide the foundation for discussing at finite temperature both the limit of infinitely stiff springs (\autoref{sec:stiff-spring limit}) and the general case of finite spring stiffness (\autoref{sec:general case}).
In the companion paper \footnote{Reference to be inserted later.}, we numerically verify the analytical results obtained here.

\section{Model}
\label{eq:model}
\subsection{Network of generalized springs}
\label{eq:model under-constrained system}
We consider a very general class of Hookean spring networks that display a strain-induced rigidity transition in the athermal limit. 

The network is embedded in $D$ spatial dimensions with periodic boundary conditions with system volume $\mathcal{V}$.
The periodic box can be sheared with shear strain variable $\gamma$, where our framework applies both to simple and pure shear. In \aref{app:PBCs and shear}, we show examples for the implementation of the periodic boundary conditions with both kinds of shear.

A network consists of $N_\mathrm{node}$ nodes whose position components combine into a $N_\mathrm{dof}=DN_\mathrm{node}$-dimensional vector $\vecn{R}$.
The nodes are connected by $N_s$ generalized springs with lengths $L_i$, labeled by $i=1,\dots,N_s$. We consider an under-constrained spring network, where $N_\mathrm{dof}>N_s$.

Each spring length depends on node positions, system size, and shear through an arbitrary function $L_i(\vecn{R},\mathcal{V},\gamma)$ that only needs to fulfill the following two criteria:
\begin{enumerate}
	\item In the vicinity of the athermal transition, the function $L_i(\vecn{R},\mathcal{V},\gamma)$ is \emph{analytic with respect to all three parameters}.
	\item The spring length $L_i$ \emph{has dimensionality $d_i$} in the following sense: For some non-negative integer $d_i$, the spring length function $L_i(\vecn{R}, \mathcal{V}, \gamma)$ is homogeneous of degree $d_i$ with respect to an isotropic rescaling. This means that for any real rescaling factor $c>0$:
	\begin{equation}
		L_i(c\vecn{R}, c^D\mathcal{V}, \gamma) = c^{d_i}L_i(\vecn{R}, \mathcal{V}, \gamma). \label{eq:L-homogeneous}
	\end{equation}
	For example, angular springs have $d_i=0$, ordinary linear springs or perimeter springs in the vertex model have $d_i=1$ (\aref{app:PBCs and shear}), and area and volume springs in the vertex model have $d_i=2$ and $d_i=3$, respectively. 
\end{enumerate}
Note that the functions $L_i$ are allowed to substantially differ among the generalized springs of the network.
For example, also models that combine 2D or 3D vertex models with spring networks are accounted for \cite{Parker2020a}.

The system energy is:
\begin{equation}
	E = \frac{1}{2}\sum_{i=1}^{N_s}{K_i\Big[L_i(\vecn{R}, \mathcal{V}, \gamma)-L_{0i}\Big]^2},
	\label{eq:energy}
\end{equation}
where we introduced spring constants $K_i$ and rest lengths $L_{0i}$.
Throughout this article, we treat all $K_i$ and $L_{0i}$ as constant and study the behavior of the $L_i$ depending on system size $\mathcal{V}$, shear $\gamma$, and temperature $T$.

\subsection{Athermal transition and isotropic strain}
\label{sec:athermal transition}
We consider a network that in the athermal limit transitions from floppy (zero shear modulus) to rigid (finite shear modulus) at some critical volume.  
For spring networks or vertex models, this is a typical scenario when the network percolates the periodic box (in the sense of connectivity percolation).
For zero shear strain, $\gamma=0$, we denote this critical volume by $\mathcal{V}^\ast$, where, consistent with our earlier findings \cite{Merkel2019,Lee2022}, we assume here that the system is floppy for $\mathcal{V}<\mathcal{V}^\ast$ and rigid for $\mathcal{V}>\mathcal{V}^\ast$.

We define the isotropic strain $\varepsilon$ with respect to the critical volume $\mathcal{V}^\ast$:
\begin{equation}
	\varepsilon := \frac{1}{D}\log{\left(\frac{\mathcal{V}}{\mathcal{V}^\ast}\right)}. \label{eq:definition-epsilon}
\end{equation}
Thus, for $\gamma=0$, the transition occurs at $\varepsilon=0$.
In other words, $\varepsilon<0$ corresponds to the athermal floppy regime with a vanishing shear modulus, and $\varepsilon>0$ corresponds to the athermal rigid regime with a finite shear modulus.

For simplicity, we assume here and in the following that there are no states of self-stress (SSS) in the floppy regime. This implies that in the floppy regime and at the transition, each spring attains its rest length, $L_i=L_{0i}$. Also, for simplicity of the argument, we assume that only a single SSS forms at the transition, which is generally true for disordered networks.

We denote the values of the dofs $\vecn{R}$ at the transition by $\vecn{R}^\ast$.
For simplicity, we assume here that $\vecn{R}^\ast$ is uniquely defined. A degeneracy in the athermal transition configuration $\vecn{R}^\ast$ could be created either by global translations or by mechanisms. Global translations do not change any of our arguments in this paper, and we deliberate more on the possibility of a $\vecn{R}^\ast$ degeneracy by mechanisms in the discussion section.

\subsection{Dimensionless quantities}
\label{sec:dimensionless quantities}
We introduce dimensionless dofs $\vecn{r}$ and dimensionless spring lengths $\ell_i$ as:
\begin{align}
	\vecn{r} := \frac{1}{\sqrt[D]{\mathcal{V}}}\vecn{R} = \frac{1}{e^{\varepsilon}\sqrt[D]{\mathcal{V}^\ast}}\vecn{R}, \label{eq:r-R} \\
	\ell_i := e^{-d_i\varepsilon}\frac{L_i(\vecn{R}, \mathcal{V}, \gamma)}{L_{0i}}. \label{eq:L-ell}
\end{align}
Using Eqs.~\eqref{eq:r-R} and \eqref{eq:L-homogeneous} with $c=1/\sqrt[D]{\mathcal{V}}=1/(e^{\varepsilon}\sqrt[D]{\mathcal{V}^\ast})$, we obtain from Eq.~\eqref{eq:L-ell}:
\begin{equation}
	\ell_i(\vecn{r},\gamma)=\frac{(\mathcal{V}^\ast)^{d_i/D}}{L_{0i}}L_i(\vecn{r}, 1, \gamma).
\end{equation}
In other words, when expressed using the dimensionless node positions $\vecn{r}$, the dimensionless spring lengths do not depend on isotropic strain.

As discussed in \autoref{sec:athermal transition}, for $\gamma=0$, the transition occurs at $\varepsilon=0$ and $\vecn{r}=\vecn{r}^\ast$, and each spring attains its rest length, $L_i=L_{0i}$. From Eq.~\eqref{eq:L-ell} thus follows that at the transition point for $\gamma=0$:
\begin{equation}
	\ell_i\big(\vecn{r}=\vecn{r}\,^\ast,\gamma=0\big)=1.	\label{eq:rest-length=1}
\end{equation}
Thus, the homogeneity property in Eq.~\eqref{eq:L-homogeneous} allows us to drop the need for a common rest lengths and equal dimensions of all springs, which was required in our earlier work \cite{Merkel2019,Lee2022}. Instead, the common rest length of all rescaled spring lengths is just one.

For convenience, we define the dimensionless weights
\begin{equation}
	w_i := \sqrt{\frac{K_iL_{0i}^2}{E_0}}, \label{eq:wi}
\end{equation}
where we have introduced the energy scale $E_0:=(\sum_i{K_iL_{0i}^2})/N_s$ such that $\sum_iw_i^2=N_s$.
This allows us to rewrite the system energy as
\begin{equation}
	E = \frac{E_0}{2}\sum_{i=1}^{N_s}{w_i^2\big(e^{d_i\varepsilon}\ell_i-1\big)^2}.
	\label{eq:energy-dimensionless}
\end{equation}
Our approach of exploiting spring length homogeneity with respect to spatial rescaling, Eq.~\eqref{eq:L-homogeneous}, thus allows us to include the overall isotropic strain explicitly into the energy functional, Eq.~\eqref{eq:energy-dimensionless-small}. This will help us to obtain our analytical results more directly than in our previous work \cite{Merkel2019,Lee2022}.

\subsection{Zero modes}
\label{sec:model - zero modes}
We close this section with some definitions of different kinds of zero modes in the system, as they will turn out to be important for its elastic behavior.

As usual \cite{Calladine1978,Lubensky2015}, for some configuration $(\vecn{r}, \varepsilon, \gamma)$, we call a vector $\vecn{z}\in\mathbb{R}^{N_\mathrm{dof}}$ \emph{infinitesimal zero mode} whenever for all $i=1,\dots,N_s$:
\begin{equation}
	\frac{\partial\ell_i(\vecn{r}, \gamma)}{\partial r_n}z_n = 0. \label{eq:inf-zero-mode}
\end{equation}
Here, $r_n$ and $z_n$ with $n=1,\dots,N_\mathrm{dof}$ are the components of the vectors $\vecn{r}$ and $\vecn{z}$, respectively, and we imply summation over repeated indices.  For a given configuration, we denote the dimension of the vector space of infinitesimal zero modes by $N_0$.

Any given infinitesimal zero mode is either trivial, it belongs to a mechanism, or it belongs to a $K$th-order zero mode.
In the absence of immobile nodes, the system shows translational invariance. We call an infinitesimal zero mode $\vecn{z}$ \emph{trivial} whenever it corresponds to one of the $D$ global translations. Conversely, in the presence of immobile nodes, there are no trivial zero modes.

A \emph{mechanism} is a finite displacement of the dofs
during which all $N_s$ springs keep their rest lengths.
More formally, we say an infinitesimal zero mode $\vecn{z}$ belongs to a mechanism whenever it is not trivial and there is a continuous function $\vecn{x}: u\mapsto \vecn{x}(u), \mathbb{R}\rightarrow\mathbb{R}^{N_\mathrm{dof}}$ with $\vecn{x}(0)=\vecn{r}$ and some range $\eta>0$ such that for all $i=1,\dots,N_s$:
\begin{align}
	\vecn{x}'(0) &=\vecn{z} &&\text{and}\\
	e^{d_i\varepsilon}\ell_i\big(\vecn{x}(u), \gamma\big) &=1 && \text{for $\vert u\vert<\eta$,} \label{eq:mechanism}
\end{align}
where the prime denotes a derivative with respect to $u$.
The second condition, Eq.~\eqref{eq:mechanism}, is equivalent to $L_i=L_{0i}$ (compare Eq.~\eqref{eq:L-ell}). For given $\varepsilon$ and $\gamma$, we denote the number of mechanisms of a system by $N_\mathrm{mech}$
\footnote{
	More precisely, we define the number of mechanisms $N_\mathrm{mech}$ of a system as the maximal integer $N_m$ for which there is a continuous function $\vecn{x}: \vecn{u}\mapsto \vecn{x}(\vecn{u}), \mathbb{R}^{N_m}\rightarrow\mathbb{R}^{N_\mathrm{dof}}$ and some $\eta>0$ such that Eq.~\eqref{eq:mechanism} holds within some open ball of radius $\eta$, i.e.\ for $\vert\vecn{u}\vert<\eta$.
}.
As discussed in \autoref{sec:athermal transition}, we focus for simplicity on a system where \emph{at the transition point}, $\vecn{r}=\vecn{r}\,^\ast$ and $\varepsilon=\gamma=0$, there are no mechanisms, $N_\mathrm{mech}^\ast=0$. Note however, that this does not preclude the existence of mechanisms away from the transition point.

A \emph{$K$th-order zero mode} does not affect the spring length to $K$th order, but to $(K+1)$th order.
More precisely, we say the infinitesimal zero mode $\vecn{z}$ belongs to a $K$th-order zero mode with $K>0$ whenever there is a $K$ times differentiable function $\vecn{x}: u\mapsto \vecn{x}(u), \mathbb{R}\rightarrow\mathbb{R}^{N_\mathrm{dof}}$  with $\vecn{x}(0)=\vecn{r}$ such that for all $i=1,\dots,N_s$:
\begin{equation}
	\vecn{x}'(0) =\vecn{z}
	\qquad \text{and}\qquad
	\left.\frac{\d^k \ell_i\big(\vecn{x}(u), \gamma\big)}{\d u^k}\right\vert_{u=0} =0
\end{equation}
for all integers $k\leq K$, \emph{while this condition is not fulfilled for $K+1$.} I.e.\ there is \emph{no} $K+1$ times differentiable function $\vecn{y}: u\mapsto \vecn{y}(u), \mathbb{R}\rightarrow\mathbb{R}^{N_\mathrm{dof}}$ with $\vecn{y}(0)=\vecn{r}$ such that for all $i=1,\dots,N_s$:
\begin{equation}
	\vecn{y}'(0) =\vecn{z}
	\qquad \text{and}\qquad
	\left.\frac{\d^{k} \ell_i\big(\vecn{y}(u), \gamma\big)}{\d u^{k}}\right\vert_{u=0} =0
\end{equation}
for all integers $k\leq K+1$. We stress that due to the last condition, a mechanism is never a $K$th-order zero mode. For the same reason, while a first-order zero mode always corresponds to an infinitesimal zero mode, the converse is not always true.  
For some configuration $(\vecn{r}, \varepsilon, \gamma)$, we denote the number of $K$th-order zero modes by $N_\mathrm{1st}, N_\mathrm{2nd}, \dots$.
Specifically, at the transition point, $\vecn{r}=\vecn{r}\,^\ast$ and $\varepsilon=\gamma=0$, we denote those numbers by $N_\mathrm{1st}^\ast, N_\mathrm{2nd}^\ast, \dots$.

While there is generally a finite number of first-order zero modes at the transition, $N_\mathrm{1st}^\ast$ (see Fig.~\ref{fig:counting examples} for examples), we expect the occurrence of $K$th-order zero modes with $K>1$ to be an exception in disordered systems. For simplicity we will focus in this article on systems without such modes at the transition point, $N_{K\mathrm{th}}^\ast=0$ for $K>1$. In \aref{app:higher-order-delta-ell-Ns}, we discuss how our results can be affected by the presence of such modes.

\section{Athermal limit (\texorpdfstring{$T\rightarrow 0$}{T=0})}
\label{sec:athermal limit}
While in our previous work we derived the elastic system properties in the athermal limit based on a minimal-length function \cite{Merkel2019,Lee2022}, we present here an alternative derivation, which will prepare our discussions in the subsequent sections.

We will minimize the energy varying the node positions, $\vecn{r}$, but keeping the strain variables $\varepsilon, \gamma$ fixed.
To this end, we first derive an explicit expression of the energy in terms of the dofs $\vec{r}$ and the strain variables $\varepsilon$ and $\gamma$ in the vicinity of the transition.

\subsection{Explicit expression of the system energy in terms of dofs and strain}
To obtain an explicit expression for the energy in Eq.~\eqref{eq:energy-dimensionless} in terms of the dofs $\vecn{r}$ and strain $\varepsilon,\gamma$  in the vicinity of the transition, we first define the deviation of the spring lengths from their transition point values (Eq.~\eqref{eq:rest-length=1}) as
\begin{equation}
	\Delta\ell_i(\vecn{r},\gamma):=\ell_i-1.
\end{equation}
This allows us to rewrite the energy, Eq.~\eqref{eq:energy-dimensionless}, to lowest order in $\varepsilon$ as:
\begin{equation}
	E = \frac{E_0}{2}\sum_{i=1}^{N_s}{w_i^2\big(\Delta\ell_i+d_i\varepsilon\big)^2}.
	\label{eq:energy-dimensionless-small}
\end{equation}
In deriving this expression from Eq.~\eqref{eq:energy-dimensionless}, we have factored out the prefactor $e^{d_i\varepsilon}$ from the parentheses. Because the parentheses scale to lowest order as $\sim\varepsilon$, the resulting prefactor of $e^{2d_i\varepsilon}$ only adds higher-order terms and is thus neglected.

To obtain an explicit expression for the $\Delta\ell_i$ in Eq.~\eqref{eq:energy-dimensionless-small} in terms of the dofs $\vecn{r}$, we expand the dimensionless spring lengths $\ell_i$ around the transition point $\vecn{r}\,^\ast$ and for small $\gamma$.
Denoting the components of $\vecn{r}$ by $r_n$ with $n=1,\dots,N_\mathrm{dof}$, we can write to second order in $\Delta r_n := r_n-r_n^\ast$ and $\gamma$:
\begin{equation}
	\begin{split}
		w_\underdot{i}\Delta\ell_\underdot{i} &= C_{in}^{(0)}\Delta r_n  + \frac{1}{2}M_{imn}^{(0)}\Delta r_m\Delta r_n \\
		&\qquad\qquad + B_i^{(1)}\gamma + C_{in}^{(1)}\Delta r_n\gamma +  \frac{1}{2}B_i^{(2)}\gamma^2. 	
	\end{split}
	\label{eq:Taylor-expansion}
\end{equation}
Here and in the following, we imply summation over repeated indices, except for indices with underdots.
In the absence of explicit sum symbols, implicit sums over the index $i$ always run over all springs ($i=1,\dots,N_s$), and implicit sums over the indices $m,n$ always run over all dofs ($m,n=1,\dots,N_\mathrm{dof}$).
On the right-hand side, $C_{in}^{(0)}\equiv C_{in}:=w_\underdot{i}(\partial\ell_\underdot{i}/\partial r_n)$ is a generalized compatibility matrix and the matrix $M_{imn}^{(0)}\equiv M_{imn}:=w_\underdot{i}(\partial^2\ell_\underdot{i}/\partial r_m\partial r_n)$ is symmetric in the last two indices.
We introduce the shear strain dependence through the coefficients $B_i^{(1)}:=w_\underdot{i}(\partial\ell_\underdot{i}/\partial\gamma)$, $C_{in}^{(1)}:=w_\underdot{i}(\partial^2\ell_\underdot{i}/\partial r_n\partial\gamma)$, and $B_i^{(2)}:=w_\underdot{i}(\partial^2\ell_\underdot{i}/\partial\gamma^2)$.
All derivatives are evaluated at the transition point, $\vecn{r}=\vecn{r}\,^\ast$, and $\gamma=0$.
Higher-order terms in $\Delta\vecn{r}$ and $\gamma$ are not important to lowest order in our final results, both here and in the following sections (\aref{app:first-order-gamma}).
There are no $\varepsilon$-dependent terms in Eq.~\eqref{eq:Taylor-expansion}, because the $\ell_i$ functions are independent of $\varepsilon$ (\autoref{sec:dimensionless quantities}).


Equations~\eqref{eq:energy-dimensionless-small} and \eqref{eq:Taylor-expansion} provide an explicit expression of the system energy in terms of dofs $\vecn{r}$, and strain variables $\varepsilon,\gamma$. 
To simplify minimizing this energy with respect to the dofs, we now rewrite $\Delta\ell_i$ and the energy $E$ in terms of eigenmodes of the compatibility matrix $C_{in}^{(0)}$. To this end, we first perform a singular-value decomposition (SVD) on it:
\begin{equation}
  C_{in}^{(0)} = \sum_{p=1}^{N_s}{U_{ip}s_pV_{np}}, \label{eq:SVD}
\end{equation}
where $U_{ip}$ and $V_{np}$ are orthogonal square matrices of dimensions $N_s\times N_s$ and $N_\mathrm{dof}\times N_\mathrm{dof}$, respectively. The $N_s$ singular values $s_p$ are sorted in decreasing order: $s_1\geq\dots\geq s_{N_s}$. Because at the transition a single SSS is created, we have $s_{N_s-1}>0$ and $s_{N_s}=0$, where $U_{iN_s}$ is the corresponding SSS.
Furthermore, all $V_{np}$ with fixed $p=N_s,\dots,N_\mathrm{dof}$ are infinitesimal zero modes.

Note that because $s_{N_s}=0$, Eq.~\eqref{eq:SVD} still holds when replacing $U_{iN_s}$ by $-U_{iN_s}$ for all $i=1,\dots,N_s$. The same is also true for the orthogonality of $U_{ip}$. In other words, the sign of the vector $(U_{iN_s})$ is not uniquely defined by Eq.~\eqref{eq:SVD}. In the following, we choose this sign such that
\begin{equation}
	\sum_{i=1}^{N_s}{U_{iN_s}d_iw_i}\geq 0\label{eq:sign_U}
\end{equation}
for later convenience.

Based on the SVD, Eq.~\eqref{eq:SVD}, we define changes of spring lengths and of degrees of freedom in the left and right eigenbases of $C_{in}^{(0)}$, respectively:
\begin{align}
  \Delta\tilde{\ell}_p &:= \sum_i{U_{ip}w_i\Delta\ell_i} &&\text{for $p=1,\dots,N_s$,}\label{eq:tildedlp}\\
  \Delta\tilde{r}_p &:= V_{np}\Delta r_n &&\text{for $p=1,\dots,N_\mathrm{dof}$.} \label{eq:tilderp}
\end{align}
As a consequence of Eq.~\eqref{eq:SVD}, $s_{N_s}=0$, and Eq.~\eqref{eq:tilderp}, all $\Delta\tilde{r}_p$ with $p=N_s,\dots,N_\mathrm{dof}$ are infinitesimal zero modes.

We can now express the Taylor expansion, Eq.~\eqref{eq:Taylor-expansion}, in terms of the eigenmodes of $C^{(0)}$. For $p=1,\dots,N_s$:
\begin{equation}
	\begin{split}
		\Delta\tilde{\ell}_p &= s_\underdot{p}\Delta\tilde{r}_\underdot{p} + \frac{1}{2}\tilde{M}_{pqr}\Delta\tilde{r}_q\Delta\tilde{r}_r\\
		&\qquad\qquad + \tilde{B}_p^{(1)}\gamma + \tilde{C}_{pq}^{(1)}\Delta\tilde{r}_q\gamma +  \frac{1}{2}\tilde{B}_p^{(2)}\gamma^2,
	\end{split}	
	\label{eq:Taylor-expansion-ebOfC}
\end{equation}
where we have also transformed $M$, $C^{(1)}$, and $B^{(1/2)}$ into the eigenbases of $C^{(0)}$:  $\tilde{M}_{pqr} := U_{ip}M_{imn}V_{mq}V_{nr}$, $\tilde{C}_{pq}^{(1)} := U_{ip}C_{in}^{(1)}V_{nq}$, and $\tilde{B}_p^{(1/2)} := U_{ip}B_i^{(1/2)}$.

We can now express the system energy, Eq.~\eqref{eq:energy-dimensionless-small}, in terms of the eigenmodes of $C^{(0)}$:
\begin{equation}
\begin{split}
  E
  &= \frac{E_0}{2}\sum_{i=1}^{N_s}{\Big(w_i\Delta\ell_i+d_iw_i\varepsilon\Big)^2} \\
  &= \frac{E_0}{2}\sum_{i,j=1}^{N_s}{\Big(w_i\Delta\ell_i+d_iw_i\varepsilon\Big)\delta_{ij}\Big(w_j\Delta\ell_j+d_jw_j\varepsilon\Big)} \\
  &= \frac{E_0}{2}\sum_{i,j,p=1}^{N_s}{\Big(w_i\Delta\ell_i+d_iw_i\varepsilon\Big)U_{ip}U_{jp}\Big(w_j\Delta\ell_j+d_jw_j\varepsilon\Big)} \\
  &= \frac{E_0}{2}\sum_{p=1}^{N_s}{\Big(\Delta\tilde\ell_p+\tilde{w}_p\varepsilon\Big)^2}.
\end{split}
\label{eq:energy-dimensionless-ebOfC}
\end{equation}
In the second step of Eq.~\eqref{eq:energy-dimensionless-ebOfC}, we introduce the Kronecker delta $\delta_{ij}$, and in the third step, we have used the orthogonality of $U$: $U_{ip}U_{jp}=\delta_{ij}$. In the fourth step we have used Eq.~\eqref{eq:tildedlp} and introduced $\tilde{w}_p:=\sum_{i=1}^{N_s}{U_{ip}d_iw_i}$.
Note that because of Eq.~\eqref{eq:sign_U}, we have $\tilde{w}_{N_s}\geq0$.

To more clearly express the dependency of the energy on the dofs $\Delta\tilde{r}_q$ and strain variables $\varepsilon,\gamma$, we define for $p=1,\dots,N_s$:
\begin{equation}
	\begin{split}
	\Lambda_p
	&:= \Delta\tilde\ell_p+\tilde{w}_p\varepsilon \\
	&= s_\underdot{p}\Delta\tilde{r}_\underdot{p} + \frac{1}{2}\tilde{M}_{pqr}\Delta\tilde{r}_q\Delta\tilde{r}_r \\ 
	&\qquad\quad +\tilde{w}_p\varepsilon + \tilde{B}_p^{(1)}\gamma + \tilde{C}_{pq}^{(1)}\Delta\tilde{r}_q\gamma +  \frac{1}{2}\tilde{B}_p^{(2)}\gamma^2.
	\end{split}
	\label{eq:Lambdap}
\end{equation}
In the second step, we used Eq.~\eqref{eq:Taylor-expansion-ebOfC}.
Inserting the definition of $\Lambda_p$ into the energy, Eq.~\eqref{eq:energy-dimensionless-ebOfC}, we obtain:
\begin{equation}
	E	= \frac{E_0}{2}\sum_{p=1}^{N_s}\Lambda_p^2.
	\label{eq:energy-dimensionless-Lambda}
\end{equation}
Thus, each $\Lambda_p$ indicates how much the corresponding left eigenmode $p$ of $C^{(0)}$ contributes to the system's energy.

\subsection{Energy minimum}
We now discuss the minimum of the energy given by Eqs.~\eqref{eq:Lambdap} and \eqref{eq:energy-dimensionless-Lambda} for fixed $\varepsilon,\gamma$ while varying the dofs $\Delta\tilde{r}_q$.
For clarity, we focus here on the case where $\tilde{B}_p^{(1)}=\tilde{C}_{pq}^{(1)}=0$ for all $p,q$. The general case of arbitrary $\tilde{B}_p^{(1)}$ and $\tilde{C}_{pq}^{(1)}$ is discussed in \aref{app:first-order-gamma}.

At $\varepsilon=\gamma=0$, the system is at its transition point $\vecn{r}=\vecn{r}\,^\ast$ (\autoref{sec:athermal transition}), which implies $\Delta\tilde{r}_q=0$ and thus, using Eq.~\eqref{eq:Lambdap}, $\Lambda_p=0$.
However, as $\varepsilon$ and $\gamma$ become nonzero, also the $\Delta\tilde{r}_q$ and $\Lambda_p$ generally become nonzero.
We provide the detailed arguments for the scaling of $\Lambda_p$ and $\Delta\tilde{r}_q$ with $\varepsilon$ and $\gamma$ in \aref{app:orders of magnitude delta r - athermal}, and provide here a brief sketch of these arguments.
Minimizing the energy in Eq.~\eqref{eq:energy-dimensionless-Lambda} can be achieved by minimizing the absolute values of all the $\Lambda_p$.
Indeed, to lowest relevant order in $\varepsilon$ and $\gamma$, the values of the $\Lambda_p$ with $p<N_s$ can be set to zero by adjusting the value of the corresponding $\Delta\tilde{r}_p$ to (compare Eq.~\eqref{eq:lambdap=0} in \aref{app:orders of magnitude delta r - athermal}):
\begin{equation}
	\begin{split}
		\Delta\tilde{r}_p = -\frac{1}{s_\underdot{p}}\Bigg(&\frac{1}{2}\sum_{q,r=N_s}^{N_\mathrm{dof}}{\tilde{M}_{\underdot{p}qr}\Delta\tilde{r}_q\Delta\tilde{r}_r} \\
		&\qquad\qquad\qquad+\tilde{w}_\underdot{p}\varepsilon +  \frac{1}{2}\tilde{B}_\underdot{p}^{(2)}\gamma^2\Bigg). \label{eq:deltarp-rough}
	\end{split}
\end{equation}
However, $\Lambda_{N_s}$ cannot be set to zero in the same way, i.e.\ with $\Delta \tilde{r}_{N_s}$ given by Eq.~\eqref{eq:deltarp-rough}, because $s_{N_s}=0$.
As a consequence, the energy is dominated by the $\Lambda_{N_s}$ contribution, $E=E_0\Lambda_{N_s}^2/2$, where, according to Eq.~\eqref{eq:Lambdap} we have:
\begin{equation}
	\Lambda_{N_s} = \frac{1}{2}\sum_{q,r=N_s}^{N_\mathrm{dof}}{\tilde{M}_{N_sqr}\Delta\tilde{r}_q\Delta\tilde{r}_r} +\tilde{w}_{N_s}\varepsilon +  \frac{1}{2}\tilde{B}_{N_s}^{(2)}\gamma^2.\label{eq:Lambda_Ns}
\end{equation}
In the sum, the contributions by the terms with $q<N_s$ or $r<N_s$ can be neglected to lowest order, since they only generate higher order terms in $\varepsilon$ and $\gamma$ (Eq.~\eqref{eq:deltarp-rough} and \aref{app:orders of magnitude delta r - athermal}).

We know that for $\gamma=0$ and $\varepsilon>0$, where the system is in the athermal rigid regime, the shear modulus is positive, and thus the energy minimum is strictly positive, $E=E_0\Lambda_{N_s}^2/2>0$.  This is only possible if the matrix $\tilde{M}_{N_sqr}$ (restricted to $q,r=N_s,\dots,N_\mathrm{dof}$) is positive semi-definite; otherwise $\Lambda_{N_s}$ could become zero.

To further discuss the energy minimum, we introduce
\begin{equation}
	b_\varepsilon := \frac{\tilde{B}_{N_s}^{(2)}}{2\tilde{w}_{N_s}}\qquad \text{if $\tilde{B}^{(1)}_p=\tilde{C}^{(1)}_{pq}=0$ for all $p,q$},\label{eq:be}
\end{equation}
so that $\Lambda_{N_s}$ becomes
\begin{equation}
	\Lambda_{N_s} = \frac{1}{2}\sum_{q,r=N_s}^{N_\mathrm{dof}}{\tilde{M}_{N_sqr}\Delta\tilde{r}_q\Delta\tilde{r}_r} +\tilde{w}_{N_s}\big[\varepsilon +  b_\varepsilon\gamma^2\big].\label{eq:Lambda_Ns_be}
\end{equation}
Because $\tilde{M}_{N_sqr}$ is positive semi-definite, we obtain that for $\varepsilon + b_\varepsilon\gamma^2>0$, the energy is minimal if $\Delta\tilde{r}_q=0$ for all $N_s\leq n\leq N_\mathrm{dof}$, and it has the value
\begin{equation}
  E = \frac{E_0\tilde{w}_{N_s}^2}{2}\Big[\varepsilon + b_\varepsilon\gamma^2\Big]^2.
\label{eq:energy-athermal}
\end{equation}
Meanwhile, for $\varepsilon + b_\varepsilon\gamma^2\leq0$, the minimal energy becomes zero since $\Lambda_{N_s}$ can become zero. Hence, $\varepsilon + b_\varepsilon\gamma^2>0$ corresponds to the athermal rigid regime and $\varepsilon + b_\varepsilon\gamma^2<0$ to the athermal floppy regime.

In the more general case where the $\tilde{B}_p^{(1)}$ and $\tilde{C}^{(1)}_{pq}$ do not vanish, the definition of $b_\varepsilon$ includes more terms, while Eq.~\eqref{eq:energy-athermal} remains the same up to a shift in $\gamma$ (\aref{app:first-order-gamma}).

Note that the matrix $\tilde{M}_{N_sqr}$ is closely linked to the Hessian of the system and the emergence of rigidity in the athermal limit (\aref{app:Hessian-M}).

\subsection{Comparison to earlier results}
The expression we published earlier for the energy of a network with only $d$-dimensional springs in the rigid regime corresponds to \cite{Merkel2019,Lee2022}:
\begin{equation}
  E = \frac{E_0N_s}{2(L_0^\ast)^2(1+a_\ell^2)}\Big[-\Delta L_0+b\gamma^2\Big]^2,
  \label{eq:energy-athermal-published}
\end{equation}
where $\Delta L_0=-dL_0^\ast\varepsilon$, and $L_0^\ast$, $a_\ell$, and $b$ are constants that depend on the network structure. We included here the prefactor of 1/2 in our energy definition as compared to \cite{Merkel2019,Lee2022}.
Comparing Eqs.~\eqref{eq:energy-athermal} and \eqref{eq:energy-athermal-published}, we find:
\begin{align}
  \tilde{w}_{N_s} &= d\left[\frac{N_s}{1+a_\ell^2}\right]^{1/2} \label{eq:tilde-e-a}\\
  b_\varepsilon &=\frac{b}{dL_0^\ast}. \label{eq:Gamma-b}
\end{align}
In addition to our earlier work \cite{Merkel2019,Lee2022}, our approach here also clarifies why the result is analytic in $\gamma$; it is inherited from the analytic nature of the spring length functions $L_i(\vecn{R},\mathcal{V},\gamma)$.

\subsection{Elastic properties}
In the rigid regime of the athermal limit, i.e.\ for $\varepsilon+b_\varepsilon\gamma^2>0$, tension $t_E=\partial E/\partial \mathcal{V}=(\partial E/\partial\varepsilon)/D\mathcal{V}$ and shear modulus $G_E=(\partial^2E/\partial\gamma^2)/\mathcal{V}$ are to lowest order in $\varepsilon$ and $\gamma$:
\begin{align}
  t_E &= \kappa_E\Big[\varepsilon+b_\varepsilon\gamma^2\Big] \label{eq:t-athermal} \\
  G_E &= 2Db_\varepsilon\kappa_E\Big[\varepsilon+3b_\varepsilon\gamma^2\Big] \label{eq:G-athermal} \\
  &= 2Db_\varepsilon t\left(1 + \frac{2b_\varepsilon\gamma^2}{\vert\varepsilon+b_\varepsilon\gamma^2\vert}\right), \label{eq:GE}
\end{align}
where we defined
\begin{align}
	\kappa_E &:= \frac{E_0\tilde{w}_{N_s}^2}{D\mathcal{V}^\ast}.
	\label{eq:kappaE}
\end{align}
The index $E$ indicates purely energetic rigidity.

\section{Stiff-spring limit (\texorpdfstring{$K_i\rightarrow \infty$}{ki=infty})}
\label{sec:stiff-spring limit}
In the limit of infinitely stiff springs but at finite temperature $T$, elasticity is created purely by entropic effects. 
For $\varepsilon+b_\varepsilon\gamma^2>0$ (rigid regime in the athermal limit), there are no configurations where all springs can attain their respective rest lengths. Thus, this regime is inaccessible in the stiff-spring limit.
Conversely, for $\varepsilon+b_\varepsilon\gamma^2\leq0$ (floppy regime in the athermal limit) there \emph{are} configurations where the springs can attain their rest lengths. In this section, we derive an expression for the accessible configurational phase space volume $\Omega$, from which we can directly derive the elastic network properties. We focus again on the limit of small $\varepsilon$ and $\gamma$.

\subsection{Configurational phase space volume}
In \aref{app:partition sum stiff-spring limit} we explicitly take the limit of the partition sum $Z$ for $K_i\rightarrow\infty$ to obtain an expression for the configurational phase space volume $\Omega$ \footnote{In the stiff-spring limit, one might naively replace each spring by an exact constraint and compute the partition sum using Lagrangian mechanics.  
However, it has been shown that such an approach generally leads to a different result than taking the partition sum $Z$ of a system with finite spring stiffnesses $K_i$ and taking the limit $K_i\rightarrow\infty$. \cite{vanKampen1984}. The latter is of course the physically more relevant approach, since there are no exact geometric constraints in nature.}.
Up to a constant factor, it is given by:
\begin{equation}
  \Omega \sim \int{\left(\prod_{n=1}^{N_\mathrm{dof}}{\d R_n}\right)\prod_{i=1}^{N_\mathrm{s}}{\delta(L_i-L_{0i})}},
	\label{eq:Omega0}
\end{equation}
where $R_n$ for $n=1,\dots,N_\mathrm{dof}$ are the components of $\vecn{R}$.
Using the dimensionless node positions and spring lengths, Eqs.~\eqref{eq:r-R} and \eqref{eq:L-ell}, this transforms into
\begin{equation}
	\Omega \sim \int{\left(\prod_{n=1}^{N_\mathrm{dof}}{\d r_n}\right)\prod_{i=1}^{N_\mathrm{s}}{\delta(\ell_i-e^{-d_i\varepsilon})}}.
	\label{eq:Omega}
\end{equation}
Here and in the following, we ignore any prefactors in $\Omega$ that are powers of $e^\varepsilon$. This is because these factors only create terms $\sim\varepsilon$ in the free energy $F_S\sim\log{\Omega}$. For small $\varepsilon$, such terms are negligible compared to the leading-order term in the free energy, which, as we will show, scales as $\sim\log{(-\varepsilon)}$ (Eq.~\eqref{eq:FS}).

We aim to obtain an analytical expression for $\Omega$ in the vicinity of the transition point, i.e.\ for $\vert\varepsilon\vert\ll1$ and $\vert\gamma\vert\ll1$.
We thus rewrite $\Omega$ as:
\begin{equation}
  \Omega \sim \int{\left(\prod_{n=1}^{N_\mathrm{dof}}{\d \Delta r_n}\right) \prod_{i=1}^{N_\mathrm{s}}{\delta\Big(w_i\Delta\ell_i+d_iw_i\varepsilon\Big)}}.
	\label{eq:Omega-delta}
\end{equation}
Here, we have included a constant factor of $w_i$ in each of the Dirac deltas, and we again ignored prefactors of $e^\varepsilon$ in $\Omega$.

To better understand which set of points $\vecn{r}$ is selected by the Dirac deltas in Eq.~\eqref{eq:Omega-delta}, we again use the eigenmodes of the compatibility matrix. To this end, we apply a rotation by $V_{np}$ to the integration variables $\Delta r_n$, and a rotation by $U_{ip}$ on the argument of the multi-dimensional Dirac delta, neither of which yields an additional factor in $\Omega$:
\begin{equation}
	\begin{split}
	  \Omega \sim 
  	&\int{\left(\prod_{q=1}^{N_\mathrm{dof}}{\d \Delta\tilde{r}_q}\right)\prod_{p=1}^{N_s}{\delta\Big(\Delta\tilde{\ell}_p+\tilde{w}_p\varepsilon\Big)}} \\
  	&= \int{\left(\prod_{q=1}^{N_\mathrm{dof}}{\d \Delta\tilde{r}_q}\right)\prod_{p=1}^{N_s}{\delta\big(\Lambda_p\big)}}.
	\end{split}\label{eq:Omega-eigenbasis}
\end{equation}
In the second step, we have used the definition of the $\Lambda_p$, Eq.~\eqref{eq:Lambdap}.
The constraints by the Dirac deltas in this equations thus imply $\Lambda_p = 0$, where close to the transition, the $\Lambda_p$ are given by Eq.~\eqref{eq:Lambdap}.

We now discuss which values of the dofs, $\Delta\tilde{r}_q$, satisfy the Dirac delta conditions in Eq.~\eqref{eq:Omega-eigenbasis}.
Here, we focus again on the case where all $\tilde{B}_p^{(1)}=\tilde{C}_{pq}^{(1)}=0$. The general case of non-zero $\tilde{B}_p^{(1)}$ and $\tilde{C}_{pq}^{(1)}$ is treated in \aref{app:first-order-gamma}.
Using similar arguments as for the athermal limit, one can show that in the $\Lambda_p$ with $p=1,\dots,N_s$, all terms of order $\sim\Delta\tilde{r}_q\Delta\tilde{r}_r$ with $q<N_s$ or $r<N_s$ can be neglected (\aref{app:orders of magnitude delta r - stiff-spring}).
Thus, we can split the integrations in Eq.~\eqref{eq:Omega-eigenbasis} as follows:
\begin{equation}
  \Omega \sim \int{\left(\prod_{q=N_s}^{N_\mathrm{dof}}{\d \Delta\tilde{r}_q}\right) \delta(\Lambda_{N_s}) 
  \prod_{p=1}^{N_s-1}{\int{\d \Delta\tilde{r}_p\;\delta(\Lambda_p)}}}.
\end{equation}
This is because to lowest order, each of the $\Delta\tilde{r}_p$ with $p<N_s$ appears only once, namely as linear term in the corresponding $\Lambda_p$.
Thus, each of the $N_s-1$ inner integrals evaluates to a constant factor of $1/s_p$.
We are thus left with:
\begin{equation}
	\Omega \sim \Omega_{N_s}(\varepsilon,\gamma),
\end{equation}
where
\begin{equation}
		\Omega_{N_s}(\varepsilon,\gamma) := \int{\left(\prod_{q=N_s}^{N_\mathrm{dof}}{\d \Delta\tilde{r}_q}\right) \delta\big(\Lambda_{N_s}\big)} \label{eq:Omega_Ns}
\end{equation}
with $\Lambda_{N_s}$ given by Eq.~\eqref{eq:Lambda_Ns_be}:
\begin{equation}
	\Lambda_{N_s} = \frac{1}{2}\sum_{q,r=N_s}^{N_\mathrm{dof}}{\tilde{M}_{N_sqr}\Delta\tilde{r}_q\Delta\tilde{r}_r} +\tilde{w}_{N_s}\big[\varepsilon +  b_\varepsilon\gamma^2\big].\label{eq:Lambda_Ns_be-repeat}
\end{equation}
To evaluate $\Omega_{N_s}(\varepsilon,\gamma)$, we first consider the symmetric matrix $\tilde{M}_{N_sqr}$ restricted to $q,r=N_s,\dots,N_\mathrm{dof}$, which appears in $\Lambda_{N_s}$. We diagonalize this $(N_\mathrm{dof}-N_s+1)\times (N_\mathrm{dof}-N_s+1)$ sub matrix:
\begin{equation}
  \tilde{M}_{N_sqr} = \sum_{k=1}^{N_\mathrm{dof}-N_s+1}{\mu_kv^k_qv^k_r}, \label{eq:diagonalization-M}
\end{equation}
where $\mu_k$ are the eigenvalues and $v^k_q$ are the associated orthonormal eigenvectors.  We sort the eigenvalues in decreasing order, $\mu_1\geq\dots\geq\mu_{N_\mathrm{dof}-N_s+1}\geq0$. All eigenvalues are non-negative, because the matrix $\tilde{M}_{N_sqr}$ is positive semi-definite.

With 
\begin{equation}
	\Delta q_k:=\sum_{q=N_s}^{N_\mathrm{dof}}{v^k_q\Delta\tilde{r}_q},\label{eq:deltaq}
\end{equation}
the expression for $\Lambda_{N_s}$, Eq.~\eqref{eq:Lambda_Ns_be-repeat}, transforms into:
\begin{equation}
  \Lambda_{N_s}
  = \frac{1}{2}\sum_{k=1}^{N_\mathrm{dof}-N_s+1}{\mu_k\Delta q_k^2} + \tilde{w}_{N_s}\big[\varepsilon + b_\varepsilon\gamma^2\big]. \label{eq:ellNs-diagonal}
\end{equation}
We now substitute $\Lambda_{N_s}$ into $\Omega_{N_s}$, Eq.~\eqref{eq:Omega_Ns}:
\begin{equation}
\begin{split}
  \Omega_{N_s} &= \int{\left(\prod_{k=1}^{N_\mathrm{dof}-N_s+1}{\d \Delta q_k}\right)}  \\
  &\times\delta\Bigg(\frac{1}{2}\sum_{k=1}^{N_\mathrm{dof}-N_s+1}{\mu_k\Delta q_k^2} + \tilde{w}_{N_s}\big[\varepsilon + b_\varepsilon\gamma^2\big]\Bigg).
  \label{eq:Omega-remaining-integrals}
\end{split}
\end{equation}
Because the $v^k_q$ represent a rotation, the differential transforms without prefactor.

\subsection{Eigenmodes of \texorpdfstring{$\tilde{M}_{N_sqr}$}{the matrix M}}
\label{sec:stiff-spring - zero modes}
To evaluate the integral in Eq.~\eqref{eq:Omega-remaining-integrals}, we first discuss how the matrix $\tilde{M}_{N_sqr}$ relates to the different kinds of zero modes in the system (see also \autoref{sec:model - zero modes}).

At the transition point, there are $N_\mathrm{dof}-N_s+1$ linearly independent infinitesimal zero modes. We consider here the zero modes $\vecn{z}_k$ with $k=1,\dots,N_\mathrm{dof}-N_s+1$ that have the components $z_{k,m}:=\sum_{q=N_s}^{N_\mathrm{dof}}{V_{mq}v_q^k}$. Insertion in Eq.~\eqref{eq:inf-zero-mode} with Eq.~\eqref{eq:SVD} confirms that they are infinitesimal zero modes. According to Eqs.~\eqref{eq:tilderp} and \eqref{eq:deltaq} these modes contribute to $\Delta\vecn{r}=\vecn{r}-\vecn{r}^\ast$ with the respective amplitudes $\Delta q_k$.

Those of the $\Delta q_k$ modes that correspond to zero eigenvalues, $\mu_k=0$, of the matrix $\tilde{M}_{N_sqr}$ do not contribute to the argument of the Dirac delta in Eq.~\eqref{eq:Omega-remaining-integrals}.
There are three possibilities for the corresponding infinitesimal zero modes $\vecn{z}_k$:
\begin{itemize}
	\item $\vecn{z}_k$ could be one of the $D$ \emph{global translations}. In this case, the integral in Eq.~\eqref{eq:Omega-remaining-integrals} leads to a constant factor.
	\item $\vecn{z}_k$ could belong to a \emph{mechanism} existing at the transition point, but here we focus on systems without such mechanisms (\autoref{sec:athermal transition}).
	However, even if we took such mechanisms into account, the integral over the corresponding manifold would yield a constant to lowest order in $\varepsilon$ and $\gamma$. Thus, any possible variation of the contribution by a mechanism with $\varepsilon$ and $\gamma$ would anyway only contribute higher-order terms.
	\item $\vecn{z}_k$ belongs to a \emph{$K$th-order zero mode with $K>1$}.
	We believe this is an exceptional case and do not consider it further in the main text, but we partly discuss it in \aref{app:higher-order-delta-ell-Ns}. 
\end{itemize}
However, none of the infinitesimal zero modes $\vecn{z}_k$ with $\mu_k=0$ can be a first-order zero mode 
\footnote{
One only needs to show that for any given $k$ with $\mu_k=0$, there exists a two-times differentiable function $u\mapsto \vecn{y}(u)$ with $\vecn{y}(0)=\vecn{r}\,^\ast$, 
$\vecn{y}'(0)=\vecn{z}_k$, and $\d^2 \ell_i(\vecn{y}(u), \gamma)/\d u^2=0$.  Indeed, such a function is given by:
\begin{equation*}
	\begin{split}
	y_m(u) &:= r_m^\ast + uV_{mq}v_q^k \\
	&\qquad\qquad - \frac{u^2}{2}\sum_{p=1}^{N_s-1}{\frac{V_{mp}}{s_p}\tilde{M}_{pqr}v_{q}^kv_{r}^k}
	\end{split}
\end{equation*}
with $m=1,\dots,N_\mathrm{dof}$.
}.

Conversely, all of the infinitesimal zero modes $\vecn{z}_k$ with $\mu_k>0$ \emph{do} contribute to the argument of the Dirac delta in Eq.~\eqref{eq:Omega-remaining-integrals}. All of them are \emph{first-order zero modes}.
This is because these modes affect $\Delta\tilde{\ell}_{N_s}$ to second order
\footnote{For any two-times differentiable function $u\mapsto\vecn{y}(u)$ with $\vecn{y}(0)=\vecn{r}\,^\ast$ and $\vecn{y}'(0)=\vecn{z}_k$, it is impossible to have $\d^2 \ell_i(\vecn{y}(u), \gamma)/\d u^2=0$ for all $i=1,\dots,N_s$, because $\d^2 \Delta\tilde\ell_{N_s}(\vecn{y}(u), \gamma)/\d u^2=\mu_k>0$.}.
These first-order zero modes dominate the scaling behavior of $\Omega_{N_s}$ close to the transition, as we will see in the following subsection.
We recall from \autoref{sec:model - zero modes} that we denote the number of first-order zero modes at the transition point by $N_\mathrm{1st}^\ast$. 

\begin{figure}[t]
	\centering
	\includegraphics[width=8.5cm]{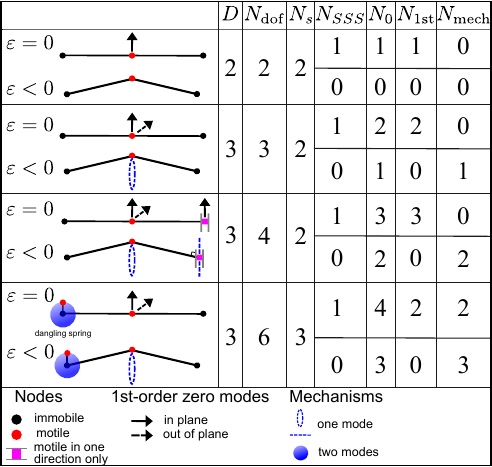}
	\caption{
		Zero mode counting for four example systems, each at the transition point ($\varepsilon=0$) and under compression ($\varepsilon<0$). In each case, we respectively indicate the number of spatial dimensions $D$, the number of degrees of freedom $N_\mathrm{dof}$, the number of springs $N_s$, the number of states of self-stress $N_{SSS}$, the number of infinitesimal zero modes $N_0$, the number of 1st-order zero modes $N_\mathrm{1st}$ (definition in \autoref{sec:model - zero modes}), and the number of mechanisms $N_\mathrm{mech}$.
		At the transition, $\varepsilon=0$, a finite number $N^\ast_\mathrm{1st}$ of 1st-order zero modes exists. Upon compression, $\varepsilon<0$, exactly $N^\ast_\mathrm{1st}-1$ of these modes turn into mechanisms. The remaining mode ceases to be an infinitesimal zero mode for $\varepsilon<0$, together with the disappearing SSS.
		While for simplicity, we focus in this article on the case where there is no mechanism at the transition point, we do not expect the existence of such mechanisms to substantially change our results. An example is shown by the fourth system, where a dangling spring creates additional mechanisms at the transition, which however do not change the physics of the system as compared to the second system.
	}
	\label{fig:counting examples}
\end{figure}
In this subsection, we so far discussed the infinitesimal zero modes at the transition point, $\varepsilon=0$.
However, the situation changes as soon as we move away from the transition point. We illustrate this  this in Fig.~\ref{fig:counting examples}, where we show a few examples for $N_\mathrm{1st}^\ast=1,\dots,3$.
As soon as we apply isotropic compression, $\varepsilon<0$ (while formally keeping $\gamma=0$), exactly $N_\mathrm{1st}^\ast-1$ of the first-order zero modes turn into mechanisms. The remaining first-order zero mode ceases to be an infinitesimal zero mode for $\varepsilon<0$. Together with this disappearing infinitesimal zero mode, the SSS that existed at $\varepsilon=0$ disappears for $\varepsilon<0$ (compare Eq.~\eqref{eq:N0}).

\subsection{Scaling of the phase space volume}
Based on our discussion in the previous section, integrating out the modes $\Delta q_k$ for $\mu_k=0$ in Eq.~\eqref{eq:Omega-remaining-integrals} only leads to factors that are to lowest order finite constants, and we obtain:
\begin{equation}
\begin{split}
  \Omega_{N_s} &\sim 
  \int{\left(\prod_{k=1}^{N_\mathrm{1st}^\ast}{\d \Delta q_k}\right)} \\
  &\qquad\quad\times\delta\left(\frac{1}{2}\sum_{k=1}^{N_\mathrm{1st}^\ast}{\mu_k\Delta q_k^2} + \tilde{w}_{N_s}\Big[\varepsilon + b_\varepsilon\gamma^2\Big]\right).
  \label{eq:Omega-remaining-integrals-2}
\end{split}
\end{equation}
The Dirac delta ensures that the integral is over the surface $\partial\mathcal{E}$ of a hyper-ellipsoid $\mathcal{E}$ in a $N_\mathrm{1st}^\ast$-dimensional space.
The surface is defined by the equation:
\begin{equation}
  \sum_{k=1}^{N_\mathrm{1st}^\ast}{\frac{\Delta q_k^2}{\rho_k^2}} = 1 \label{eq:ellipsoid-surface}
\end{equation}
with half axes $\rho_k:=(-2\tilde{w}_{N_s}[\varepsilon +b_\varepsilon\gamma^2]/\mu_k)^{1/2}\sim [-(\varepsilon +b_\varepsilon\gamma^2)]^{1/2}$.
In particular, for $\gamma=0$, the hyper-ellipsoid axes scale as $\rho_k\sim\sqrt{-\varepsilon}$.
Thus, consistent with the previous section (\autoref{sec:stiff-spring - zero modes}), as soon as compression is applied, $\varepsilon<0$, there are $N_\mathrm{1st}^\ast-1$ mechanisms, which correspond to the surface manifold of the hyper-ellipsoid.
Meanwhile, movement perpendicular to the ellipsoid surface is not allowed, corresponding to the first-order zero mode that ceases to be an infinitesimal zero mode for $\varepsilon<0$.

Using standard transformation rules of the Dirac delta, we get:
\begin{equation}
  \Omega_{N_s} \sim 
  \int_{\partial\mathcal{E}}{\d S\;\left\vert\frac{\partial \Lambda_{N_s}}{\partial \Delta\vecn{q}}\right\vert^{-1}},
  \label{eq:Omega-remaining-integral-ellipsoid-surface}
\end{equation}
where $\d S$ is an integration element of the hyper-ellipsoid surface $\partial\mathcal{E}$, and
\begin{equation}
\begin{split}
  \left\vert\frac{\partial \Lambda_{N_s}}{\partial\Delta \vecn{q}}\right\vert^2
  &= \sum_{k=1}^{N_\mathrm{1st}^\ast}{\left(\frac{\partial \Lambda_{N_s}}{\partial \Delta q_k}\right)^2} \\
  &= \sum_{k=1}^{N_\mathrm{1st}^\ast}{\mu_k^2\Delta q_k^2}.
\end{split}
\end{equation}
With $\mu_\mathrm{max}:=\mu_1$ and $\mu_\mathrm{min}:=\mu_{N_\mathrm{1st}^\ast}>0$, we thus have on the ellipsoid surface, using Eq.~\eqref{eq:ellipsoid-surface}:
\begin{equation}
  -\mu_\mathrm{min}\big[\varepsilon +b_\varepsilon\gamma^2\big]
  \leq \frac{1}{2\tilde{w}_{N_s}}\left\vert\frac{\partial \Lambda_{N_s}}{\partial\Delta\vecn{q}}\right\vert^2 
  \leq -\mu_\mathrm{max}\big[\varepsilon +b_\varepsilon\gamma^2\big].
\end{equation}
Thus, $\left\vert\partial \Lambda_{N_s}/\partial \Delta\vecn{q}\right\vert \sim [-(\varepsilon +b_\varepsilon\gamma^2)]^{1/2}$.
As a consequence, we obtain that the integral in Eq.~\eqref{eq:Omega-remaining-integral-ellipsoid-surface} scales as the hyper-surface area of the ellipsoid divided by $[-(\varepsilon+b_\varepsilon\gamma^2)]^{1/2}$:
\begin{equation}
	\Omega_{N_s}(\varepsilon,\gamma) 
	\sim 
\Big[-(\varepsilon+b_\varepsilon\gamma^2)\Big]^{(N_\mathrm{1st}^\ast-2)/2}. \\
\label{eq:Omega-Ns-final-result}
\end{equation}
Hence, $\Omega\sim\Omega_{N_s}$ scales as a power law with the distance to the transition point as long as $N_\mathrm{1st}^\ast>2$.
In \aref{app:higher-order-delta-ell-Ns}, we prove this result in a more general context, where we show that the exponent in Eq.~\eqref{eq:Omega-Ns-final-result} can change if we allow for $K$th-order zero modes with $K>1$ at the transition.

In \aref{app:first-order-gamma}, we discuss the more general case where the $\tilde{B}_p^{(1)}$ and $\tilde{C}_{pq}^{(1)}$ do not vanish. The definition of $b_\varepsilon$ changes like in the athermal case, while the result Eq.~\eqref{eq:Omega-Ns-final-result} remains the same up to an additional term $B_{N_s}^{(1)}\gamma$, which can be removed through redefining $\gamma$.

\subsection{Free energy, tension, and shear modulus}
The free energy of the system in the stiff-spring limit, $F_S$, is up to a constant:
\begin{equation}
\begin{split}
  F_S &= -\frac{1}{\beta}\log\Omega \\
  &= -\frac{N_\mathrm{1st}^\ast-2}{2\beta}\log{\Big[-(\varepsilon+b_\varepsilon\gamma^2)\Big]},
\end{split}\label{eq:FS}
\end{equation}
where $\beta:=k_BT$ with $k_B$ being the Boltzmann constant.
The index $S$ stresses that $F_S$ results entirely from entropic effects.

We then obtain for the tension $t_S=(\partial F_S/\partial\varepsilon)/D\mathcal{V}$ and the shear modulus $G_S=(\partial^2 F_S/\partial\gamma^2)/\mathcal{V}$, to lowest order:
\begin{align}
  t_S &= -\frac{\kappa_ST}{\varepsilon+b_\varepsilon\gamma^2} \label{eq:tS}\\
  G_S 
  &= -2Db_\varepsilon\kappa_ST\frac{\varepsilon-b_\varepsilon\gamma^2}{(\varepsilon+b_\varepsilon\gamma^2)^2} \\
  &= 2Db_\varepsilon t\left(1 + \frac{2b_\varepsilon\gamma^2}{\vert\varepsilon+b_\varepsilon\gamma^2\vert}\right) \label{eq:GS}
\end{align}
with
\begin{equation}
  \kappa_S := \frac{k_B(N_\mathrm{1st}^\ast-2)}{2D\mathcal{V}^\ast}. \label{eq:kappaS}
\end{equation}
Note that only the regime with $\varepsilon+b_\varepsilon\gamma^2<0$ is accessible here, which corresponds to the athermal floppy regime.

\section{General case}
\label{sec:general case}
\subsection{Partition sum}
To study how entropic and energetic elasticity interact, we evaluate the partition sum of the system
\begin{equation}
  Z = \int{\left(\prod_{n=1}^{N_\mathrm{dof}}{\d R_n}\right) e^{-\beta E}}. \label{eq:Z}
\end{equation}
Transforming the dofs $R_n$ to the dimensionless eigenmodes of the compatibility matrix, $\Delta\tilde{r}_q$, we have:
\begin{equation}
  Z \sim \int{\left(\prod_{q=1}^{N_\mathrm{dof}}{\d\Delta\tilde{r}_q}\right) e^{-\beta E}}, \label{eq:Z-dimensionless}
\end{equation}
where we again ignore prefactors proportional to $e^\varepsilon$.

We evaluate the partition sum for small $\varepsilon$, $\gamma$, and $T$.
Because $\varepsilon\ll1$, we can use the energy expression from Eq.~\eqref{eq:energy-dimensionless-Lambda}:
\begin{equation}
  E = \frac{E_0}{2}\sum_{p=1}^{N_s}{\Lambda_p^2}.
\end{equation}
Here we discuss the case where all $\tilde{B}_p^{(1)}=\tilde{C}_{pq}^{(1)}=0$, while the more general case of arbitrary $\tilde{B}_p^{(1)}$ and $\tilde{C}_{pq}^{(1)}$ is discussed in \aref{app:first-order-gamma}.

We focus on small temperatures, $k_BT\lesssim E_0[\varepsilon+b_\varepsilon\gamma^2]^2$, such that states with higher energies $E$ receive an exponentially smaller Boltzmann weight.  
Using similar arguments as in the athermal and stiff-spring limits, we find again that in all $\Lambda_q$ with $q\leq N_s$ all the $\Delta\tilde{r}_p$ with $p<N_s$ can be neglected in the terms $\sim\Delta\tilde{r}_q\Delta\tilde{r}_r$ (details in \aref{app:orders of magnitude delta r - general}). Thus, to lowest order each of the $\Delta\tilde{r}_p$ with $p<N_s$ appears only once, namely in the linear term of the respective $\Lambda_p$.
As a consequence, the integrals in the partition sum in Eq.~\eqref{eq:Z-dimensionless} can be rearranged as follows:
\begin{equation}
  Z \sim \int{\left(\prod_{q=N_s}^{N_\mathrm{dof}}{\d\Delta \tilde{r}_q}\right)\;e^{-\frac{\beta E_0\Lambda_{N_s}^2}{2}}}\prod_{p=1}^{N_s-1}{\int{\d\Delta\tilde{r}_p}\; e^{-\frac{\beta E_0\Lambda_p^2}{2}}}. \label{eq:Z-decomposed}
\end{equation}
The innermost first $N_s-1$ integrals are Gaussian integrals, each of which evaluates to the constant $Z_p=\sqrt{2\pi/\beta E_0s_p^2}$.

In a system where each of the nodes also has a mass $m$, there is an additional constant prefactor in the partition sum in Eq.~\eqref{eq:Z}, which arises from the kinetic energy of the nodes \cite{Tuckerman2010}. This prefactor is given by $\lambda^{-DN_\mathrm{node}}$, where $\lambda$ is the thermal wavelength $\lambda:=\sqrt{\beta h^2/2\pi m}$ with Planck constant $h$.
When combining each of the $Z_p$ for $p<N_s$ with one of the thermal wavelength factors, one obtains the partition sum of a harmonic oscillator, $Z_p/\lambda=2\pi/\beta h\omega_p$, with eigen frequency $\omega_p=s_p\sqrt{E_0/m}$.
This means that each of the nonzero eigen modes of the Hessian at the transition point just creates a harmonic oscillator, like for isostatic and over-constrained systems. 
However, for under-constrained systems, we additionally have the last integral, which stems from the SSS that forms at the transition:
\begin{equation}
	Z_{N_s} := \int{\left(\prod_{q=N_s}^{N_\mathrm{dof}}{\d\Delta \tilde{r}_q}\right)\;e^{-\frac{\beta E_0\Lambda_{N_s}^2}{2}}}. \label{eq:ZNs}
\end{equation}
Neglecting the factors $Z_p$ with $p<N_s$, which solely depend on temperature, but not on strain, we have $Z\sim Z_{N_s}$.

\subsection{Saddle point approximation}
To simplify the remaining integral $Z_{N_s}$ in Eq.~\eqref{eq:ZNs}, we introduce in its integrand an additional integral over some variable $\varepsilon_S$ that trivially evaluates to one:
\begin{equation}
\begin{split}
  Z_{N_s} &= \int{\left(\prod_{q=N_s}^{N_\mathrm{dof}}{\d\Delta \tilde{r}_q}\right)\;e^{-\frac{\beta E_0\Lambda_{N_s}^2}{2}}} \\
  &\qquad\times \tilde{w}_{N_s}\int_{-\infty}^{\infty}{\d\varepsilon_S\;\delta\Big(\Lambda_{N_s}+\tilde{w}_{N_s}[\varepsilon_S-\varepsilon]\Big)}.
  \label{eq:GC-SSS-1}
\end{split}
\end{equation}
After exchanging the order of integration and substituting $\Omega_{N_s}$, defined in Eq.~\eqref{eq:Omega_Ns}, we have:
\begin{equation}
	\begin{split}
		Z_{N_s} &= \tilde{w}_{N_s}\int_{-\infty}^{\infty}{\d\varepsilon_S\; \Omega_{N_s}(\varepsilon_S,\gamma)e^{-\beta E_{N_s}(\varepsilon-\varepsilon_S)}}, \label{eq:GC-SSS-2}
	\end{split}
\end{equation}
with:
\begin{equation}
	\begin{split}
	E_{N_s}(\varepsilon-\varepsilon_S) &:= \frac{E_0\tilde{w}_{N_s}^2}{2}\big(\varepsilon-\varepsilon_S\big)^2 \\
	&= \frac{D\mathcal{V}^\ast\kappa_E}{2}\big(\varepsilon-\varepsilon_S\big)^2.  
	\end{split}
\end{equation}
Here, we used the definition for $\kappa_E$ in Eq.~\eqref{eq:kappaE}.
From the discussion in the previous section, we know that for $\varepsilon_S+b_\varepsilon\gamma^2>0$, no configurations are possible, i.e.\ $\Omega_{N_s}(\varepsilon_S,\gamma)=0$. Meanwhile, for $\varepsilon_S+b_\varepsilon\gamma^2\leq 0$, we have from  Eq.~\eqref{eq:Omega-Ns-final-result} that:
\begin{equation}
	\Omega_{N_s}(\varepsilon_S,\gamma) \sim \Big[-(\varepsilon_S+b_\varepsilon\gamma^2)\Big]^{(N_\mathrm{1st}^\ast-2)/2}.
\end{equation}
These transformations identify $\varepsilon_S$ as the entropic strain discussed in the companion paper \cite{Note1}.

We use a saddle point approximation to simplify the integral in Eq.~\eqref{eq:GC-SSS-2}:
\begin{equation}
	Z_{N_s} = \tilde{w}_{N_s} \int_{-\infty}^{\infty}{\d\varepsilon_S\; e^{-\beta\bar{F}}}, \label{eq:GC-SSS-3}
\end{equation}
where $\bar{F}$ is the free energy of the system for imposed $\varepsilon_S$:
\begin{equation}
	\begin{split}
		\bar{F}(\varepsilon_S;\varepsilon,\gamma) 
		&:= E_{N_s}(\varepsilon-\varepsilon_S) -\frac{1}{\beta}\log{\Omega_{N_s}(\varepsilon_S,\gamma)} \\
		&= D\mathcal{V}^\ast\bigg( \frac{\kappa_E}{2}\big(\varepsilon-\varepsilon_S\big)^2\\
		&\qquad\qquad -\kappa_ST\log{\Big[-(\varepsilon_S +b_\varepsilon\gamma^2)\Big]}\bigg)
	\end{split}
\end{equation}
On the second line, we used the definition of $\kappa_S$, Eq.~\eqref{eq:kappaS}, and ignored a constant offset.

For small temperatures, we can apply the saddle point approximation, where we Taylor expand $\bar{F}(\varepsilon_S)$ to second order around its minimum at $\varepsilon_S=\hat\varepsilon_S$, which results in:
\begin{equation}
	Z_{N_s} \sim \int_{-\infty}^{\infty}{\d\varepsilon_S\;e^{-\beta\bar{F}}} = e^{-\beta\bar{F}(\hat\varepsilon_S)}\sqrt{\frac{2\pi}{\beta \bar{F}''(\hat\varepsilon_S)}},
\end{equation}
where $\bar{F}'':=\partial^2 F/\partial\varepsilon_S^2$:
\begin{equation}
	\begin{split}
		\bar{F}''(\hat\varepsilon_S)
		&= D\mathcal{V}^\ast\Bigg(\kappa_E + \frac{\kappa_ST}{[\hat\varepsilon_S +b_\varepsilon\gamma^2]^2} \Bigg).
	\end{split}
\end{equation}
To find the minimum $\hat\varepsilon_S$, we transform $\partial\bar{F}/\partial\varepsilon_S=0$, and obtain:
\begin{equation}
	\hat\varepsilon_S + b_\varepsilon\gamma^2 = \frac{\varepsilon + b_\varepsilon\gamma^2}{2} - \sqrt{1+\theta}\left\vert\frac{\varepsilon + b_\varepsilon\gamma^2}{2}\right\vert,
	\label{eq:hat-eS}
\end{equation}
with 
\begin{equation}
	\begin{split}
			\theta &:= \frac{4\kappa_ST}{\kappa_E(\varepsilon+b_\varepsilon\gamma^2)^2} \\
			&= \frac{2(N_\mathrm{1st}^\ast-2)k_BT}{E_0\tilde{w}_{N_s}^2(\varepsilon+b_\varepsilon\gamma^2)^2}.
	\end{split}
\label{eq:theta}
\end{equation}
Note that for $\gamma=0$, the equation $\partial\bar{F}(\hat\varepsilon_S)/\partial\varepsilon_S=0$, defining the minimum $\hat\varepsilon_S$ of $\bar{F}(\varepsilon_S)=E(\varepsilon-\varepsilon_S)+F_S(\varepsilon_S)$, corresponds to equating energetic and entropic tension, like in our intuitive explanation in the companion paper \cite{Note1}.

\subsection{Free energy, tension, and shear modulus}
Up to terms that depend only on temperature and up to higher-order terms in the strain variables, the free energy of the system, $F = -(\log{Z})/\beta$, is:
\begin{equation}
	F = \bar{F}(\hat\varepsilon_S) + \frac{1}{2\beta}\log{\bar{F}''(\hat\varepsilon_S)}. \label{eq:general-F}
\end{equation}
After some transformations, we find for the isotropic tension $t=(\partial F/\partial \varepsilon)/D\mathcal{V}$ and the shear modulus $G=(\partial^2 F/\partial\gamma^2)/\mathcal{V}$, to lowest order in $\varepsilon$ and $\gamma$ (\aref{app:derivatives general case}):
\begin{equation}
	t = \kappa_E\big(\varepsilon-\hat\varepsilon_S\big)\left[1+\frac{1}{2(N_\mathrm{1st}^\ast-2)}\;\frac{\theta}{1+\theta}\right]
\end{equation}
and
\begin{equation}
\begin{split}
	G &= 2Db_\varepsilon\kappa_E\big(\varepsilon-\hat\varepsilon_S\big)\Bigg[1+\frac{2b_\varepsilon\gamma^2}{\vert\varepsilon+b_\varepsilon\gamma^2\vert\sqrt{1+\theta}} \\
	&\quad+\frac{1}{2(N_\mathrm{1st}^\ast-2)}\;\frac{\theta}{1+\theta}\Bigg(1
	+\frac{2b_\varepsilon\gamma^2}{\vert\varepsilon+b_\varepsilon\gamma^2\vert\sqrt{1+\theta}} \\
	&\qquad\qquad\qquad\qquad\qquad\;\; -\frac{4b_\varepsilon\gamma^2}{(\varepsilon+b_\varepsilon\gamma^2)(1+\theta)}
	\Bigg)\Bigg]. \label{eq:G-general}
\end{split}
\end{equation}
Note that the last terms in both $t$ and $G$ scale as $\sim(N_\mathrm{1st}^\ast-2)^{-1}$.  These terms stem from the term $\sim\log{\bar{F}''}$ in the free energy. For $N_\mathrm{1st}^\ast\gg 1$, these contributions can be neglected, and we have:
\begin{align}
	t &= \kappa_E\big(\varepsilon-\hat\varepsilon_S\big) \label{eq:final-t}\\
	G 
	&= 2Db_\varepsilon\kappa_E\big(\varepsilon-\hat\varepsilon_S\big)\Bigg[1+\frac{2b_\varepsilon\gamma^2}{\vert\varepsilon+b_\varepsilon\gamma^2\vert\sqrt{1+\theta}}\Bigg] \\
	&= 2Db_\varepsilon t\Bigg[1+\frac{2b_\varepsilon\gamma^2}{\vert\varepsilon+b_\varepsilon\gamma^2\vert\sqrt{1+\theta}}\Bigg]. \label{eq:final-G}
\end{align}
In these equations, the value of $\hat\varepsilon_S$ can be obtained using Eq.~\eqref{eq:hat-eS}.

\subsection{Limits}
\label{sec:general case - limits}
\subsubsection{Small temperature\texorpdfstring{, $\theta\ll 1$}{}}
For $\theta\ll1$, Eq.~\eqref{eq:hat-eS} becomes:
\begin{equation}
\begin{split}
 	\hat\varepsilon_S + b_\varepsilon\gamma^2  
  &= \frac{\varepsilon + b_\varepsilon\gamma^2}{2} - \left(1 +\frac{\theta}{2}\right)\left\vert\frac{\varepsilon + b_\varepsilon\gamma^2}{2}\right\vert \\
  &= (\varepsilon + b_\varepsilon\gamma^2)\times\begin{cases}
    1+\theta/4
    &\text{for $\varepsilon + b_\varepsilon\gamma^2<0$,} \\
    -\theta/4
    &\text{for $\varepsilon + b_\varepsilon\gamma^2>0$.}
  \end{cases}
\end{split}
\end{equation}
Thus, to lowest order in $\theta$:
\begin{equation}
	\varepsilon - \hat\varepsilon_S
	= (\varepsilon + b_\varepsilon\gamma^2)\times\begin{cases}
			-\theta/4
			&\text{for $\varepsilon + b_\varepsilon\gamma^2<0$,} \\
			1
			&\text{for $\varepsilon + b_\varepsilon\gamma^2>0$,}
		\end{cases}
\end{equation}
and we recover the purely entropic and energetic tensions and shear moduli, respectively (compare Eqs.~\eqref{eq:t-athermal}, \eqref{eq:GE}, \eqref{eq:tS}, and \eqref{eq:GS}):
\begin{align}
  t & = \begin{cases}
    t_S
    &\text{for $\varepsilon + b_\varepsilon\gamma^2<0$,} \\
    t_E
    &\text{for $\varepsilon + b_\varepsilon\gamma^2>0$,}
  \end{cases}\\
  G &= \begin{cases}
	G_S
	&\text{for $\varepsilon + b_\varepsilon\gamma^2<0$,} \\
	G_E
	&\text{for $\varepsilon + b_\varepsilon\gamma^2>0$.}
\end{cases}
\end{align}
Note that for $\varepsilon + b_\varepsilon\gamma^2$ close to zero, we are not in the limit $\theta\ll 1$ any more (see Eq.~\eqref{eq:theta}).

\subsubsection{Small strain\texorpdfstring{, $\theta\gg 1$}{}}
For $\theta\gg1$, Eq.~\eqref{eq:hat-eS} becomes:
\begin{equation}
	\hat\varepsilon_S + b_\varepsilon\gamma^2  
		= -\sqrt{\frac{\kappa_ST}{\kappa_E}} + \frac{\varepsilon + b_\varepsilon\gamma^2}{2}
\end{equation}
And thus:
\begin{equation}
	\varepsilon - \hat\varepsilon_S
	= \sqrt{\frac{\kappa_ST}{\kappa_E}} + \frac{\varepsilon + b_\varepsilon\gamma^2}{2}.
\end{equation}
We then find:
\begin{equation}
	t = \frac{2N_\mathrm{1st}^\ast-3}{2N_\mathrm{1st}^\ast-4}\left(\sqrt{\kappa_E\kappa_ST} + \kappa_E\frac{\varepsilon + b_\varepsilon\gamma^2}{2}\right),
\end{equation}
and for the shear modulus:
\begin{equation}
\begin{split}
	G &= Db_\varepsilon\bigg[2\sqrt{\kappa_E\kappa_ST} + \kappa_E(\varepsilon + b_\varepsilon\gamma^2)\bigg] \\
	&\qquad\times\Bigg[\frac{2N_\mathrm{1st}^\ast-3}{2N_\mathrm{1st}^\ast-4}\Bigg(1
	+b_\varepsilon\gamma^2\sqrt{\frac{\kappa_E}{\kappa_ST}}\Bigg) \\
	&\qquad\qquad\qquad\qquad -\frac{b_\varepsilon\gamma^2}{2N_\mathrm{1st}^\ast-4}\;\frac{\kappa_E(\varepsilon+b_\varepsilon\gamma^2)}{\kappa_ST}\Bigg)\Bigg].
\end{split}
\end{equation}
For $N_\mathrm{1st}^\ast\gg 1$ both expressions simplify to:
\begin{align}
	t &= \sqrt{\kappa_E\kappa_ST} + \kappa_E\frac{\varepsilon + b_\varepsilon\gamma^2}{2} \\
	G &= 2Db_\varepsilon t\Bigg(1
		+b_\varepsilon\gamma^2\sqrt{\frac{\kappa_E}{\kappa_ST}}\Bigg).
\end{align}
Note that we assumed in our derivation for the general case that $k_BT\lesssim E_0[\varepsilon+b_\varepsilon\gamma^2]^2$. Using the definitions of $\theta$ in Eq.~\eqref{eq:theta}, of $\kappa_E$ in Eq.~\eqref{eq:kappaE}, and of $\kappa_S$ in Eq.~\eqref{eq:kappaS}, this condition corresponds to $\theta\lesssim2(N_\mathrm{1st}^\ast-2)/\tilde{w}_{N_s}^2$. Thus, this limit corresponds to $1\ll\theta\lesssim2(N_\mathrm{1st}^\ast-2)/\tilde{w}_{N_s}^2$, which can be fulfilled for $N_\mathrm{1st}^\ast\gg 1$. 
Note that in our numerical results for randomly-cut triangular networks, we also found that our prediction was accurate for finite temperatures even for vanishing strain, $\varepsilon+b_\varepsilon\gamma^2\rightarrow 0$ \cite{Note1}.

\section{Discussion}
\label{sec:discussion}
In this article, we developed a theory for thermal, under-constrained systems with fixed connectivity. 
We provide analytical expressions for the partition sum $Z$, free energy $F$, and elastic material properties such as tension $t$ and shear modulus $G$ close to the athermal rigidity transition depending on isotropic strain $\varepsilon$, shear strain $\gamma$, and temperature $T$.
The only three parameters appearing in our theory are energetic rigidity $\kappa_E$, entropic rigidity $\kappa_S$, and an interaction parameter between isotropic and shear strain, which we denote by $b_\varepsilon$.  These parameters depend on the microscopic structure of the system through Eqs.~\eqref{eq:kappaE}, \eqref{eq:kappaS}, and \eqref{eq:be} (or, more generally, Eq.~\eqref{eq:be-with-linear-order-terms}), respectively.

We first discussed the limit of zero temperature (athermal limit), which is dominated by energetic rigidity, and the limit of infinitely stiff springs, which is dominated by entropic rigidity.
In the athermal limit, we analytically derived the elastic properties before \cite{Merkel2019,Lee2022}, where the system is floppy for $\varepsilon+b_\varepsilon\gamma^2\leq0$ and rigid for $\varepsilon+b_\varepsilon\gamma^2>0$.
For the stiff-spring limit, we showed that the accessible phase space volume corresponds to the surface of a hyper-ellipsoid, whose dimension is given by the number of first-order zero modes of the system, $N_\mathrm{1st}^\ast$ (\autoref{sec:model - zero modes}). This ultimately leads to a free energy that scales logarithmically with the negative combined strain $-(\varepsilon+b_\varepsilon\gamma^2)$.
In both limits, the mechanical properties are inherently linked to the properties of the SSS that is created at the transition.
Finally, we derive the partition sum for the general case, where energetic and entropic rigidity are coupled only through the SSS.

In the regime with $\epsilon < 0$, previous numerical work has reported a shear modulus scaling of $G \sim T^{0.8}$
\cite{Dennison2013,Arzash2023}, while still reporting a linear scaling of the isotropic tension, $t\sim T^1$ \cite{Arzash2023}.
This partly contradicts our results here, where in the regime with $\varepsilon<0$ and small temperatures ($\theta \ll 1$), we obtain the scaling relations $G \sim t \sim T^1$.
We do not know the precise reason for the deviation of the shear modulus scaling, but we see the following possibilities.
First, all results obtained here are based on the Taylor expansion in Eq.~\eqref{eq:Taylor-expansion}, which holds for small deviations of the node positions from the transition point configuration, and for small shear strain $\gamma$.
A deviation of shear modulus scaling but not of tension scaling could thus be explained by too high values of the shear strain $\gamma$ for the configurations studied in Ref.~\cite{Arzash2023}, such that terms of higher than quadratic order in $\gamma$ become important in Eq.~\eqref{eq:Taylor-expansion}.

Second, Refs.~\cite{Dennison2013,Arzash2023} seem to treat shear strain differently than we do here. 
While we shear stabilize our configurations before the thermal simulations \cite{Note1}, the authors of Ref.~\cite{Dennison2013} do not seem to have done this, which can also affect the apparent shear modulus scaling as we have shown before \cite{Lee2022}.

Third, in our simulations \cite{Note1}, we found that the numerically observed value of the shear modulus $G$ can be more sensitive to insufficient thermal equilibration than the system energy or the components of the stress tensor are.  
Moreover, for the same number of simulations steps, insufficient equilibration would more strongly affect the smaller temperatures. This might be consistent with the results in Ref.~\cite{Dennison2013}, where a scaling of $G \sim T^{0.8}$ was observed for small temperatures, while $G \sim T^1$ was observed for larger temperatures.

Finally, Refs.~\cite{Dennison2013,Arzash2023} suggested that the shear modulus scaling exponent different from one was due to purely entropic effects in combination with the disordered nature of the networks. However, in our approach, we fully take the disordered network structure into account without any kind of mean-field-like assumption, and we obtain an exponent of exactly 1.
Indeed, in the infinitely-stiff-spring limit, where elasticity is of purely entropic nature, an exponent of exactly 1 is expected purely on dimensional grounds since the temperature $T$ provides the only energy scale.

While not explicitly studied here, our results have immediate consequences with respect to the scaling with network connectivity $z$.
For $\gamma=0$, earlier work by Zhang and Mao using EMT indicated three scaling regimes for the shear modulus $G$ of randomly-cut triangular networks \cite{Zhang2016a}. Remarkably, these scaling regimes are very similar to what we report in \autoref{sec:general case - limits}, where up to a prefactor, isotropic strain $\varepsilon$ is replaced by the relative connectivity $\Delta z=z-z_c$, with $z_c$ being the connectivity at isostaticity \cite{Zhang2016a}.
To understand how this could be, we discuss the $\Delta z$ scaling that results from our findings by separately discussing the $\Delta z$ scaling for $\kappa_S$ and $\kappa_E$.
Eq.~\eqref{eq:kappaS} predicts that $\kappa_S$ essentially only depends on the number of first-order zero modes $N_\mathrm{1st}^\ast$.
Thus, if for large systems $N_\mathrm{1st}^\ast$ scales to dominating order linearly with $\Delta z=-4(N_\mathrm{dof}-N_s)/N_\mathrm{dof}$, then this implies $\kappa_S\sim\vert\Delta z\vert^1$.
Meanwhile, how energetic elasticity $\kappa_E$ scales with connectivity $\Delta z$ depends on the class of the network \cite{Lee2022}. For instance, for $\vert\Delta z\vert\ll 1$ our results for the $a_\ell$ scaling in Refs.~\cite{Merkel2019,Lee2022} imply that $\kappa_E\sim\vert\Delta z\vert^2$ for Delaunay networks, $\kappa_E\sim\vert\Delta z\vert^1$ for randomly-cut, packing-derived networks, and $\kappa_E\sim\vert\Delta z\vert^0$ for phantom triangular networks.
Because of Eq.~\eqref{eq:final-t}, these scaling relations of $\kappa_S$ and $\kappa_E$ with $\Delta z$ determine the scaling of tension $t$.
However, for the scaling of the shear modulus $G$, Eq.~\eqref{eq:final-G}, the scalings of $b_\varepsilon\kappa_S$ and $b_\varepsilon\kappa_E$ are the relevant ones, and thus the scaling of $b_\varepsilon\sim b$ needs to be taken into account.
Based on earlier results for the scaling of $b$ \cite{Merkel2019,Lee2022}, we have $b_\varepsilon\kappa_S\sim\vert\Delta z\vert^{-1}$ and $b_\varepsilon\kappa_E\sim\vert\Delta z\vert^0$ for Delaunay networks, $b_\varepsilon\kappa_S\sim\vert\Delta z\vert^{0}$ and $b_\varepsilon\kappa_E\sim\vert\Delta z\vert^0$ for randomly-cut, packing-derived networks, and $b_\varepsilon\kappa_S\sim\vert\Delta z\vert^{0.5}$ and $b_\varepsilon\kappa_E\sim\vert\Delta z\vert^{-0.5}$ for phantom triangular networks.
Taken together, this suggests that the scaling of the elastic properties with $\Delta z$ is expected to depend on the class of network studied.
For instance, the results by Zhang and Mao \cite{Zhang2016a} can be explained from the perspective of our results if the randomly-cut triangular networks have $b_\varepsilon\kappa_S\sim\vert\Delta z\vert^{0}$ and $b_\varepsilon\kappa_E\sim\vert\Delta z\vert^0$ like randomly-cut, packing-derived networks.  In this case, the scaling is entirely determined by the scaling of the strain of the network state after network initialization, $\varepsilon_\mathrm{ini}$, with connectivity $\Delta z$, which is generally linear to lowest order: $\varepsilon_\mathrm{ini}\sim\Delta z$ (remember that we define isotropic strain  here with respect to the athermal transition point) \cite{Merkel2019,Lee2022}.  This would be a simple explanation for why the shear modulus of Zhang and Mao has the same form as our results when replacing $\varepsilon$ by $\Delta z$.
We leave a more thorough numerical test of all these predicitons for the scaling with $\Delta z$, and a study of their consequences, for future work.

We introduced under-constrained systems as systems that have $N_\mathrm{dof}>N_s$, and our results apply to a broad subset of such systems. In particular, we focused on systems that fulfill the following criteria:
\begin{itemize}
	\item \emph{There is no SSS in the floppy regime.} 
	Lifting the restriction on the number of SSS in the floppy regime probably does not change our results. 
	\item \emph{Only a single SSS forms at the transition.} 
	It is well-known that at least one SSS forms at the transition \cite{Lubensky2015}.
	Moreover, situations where more than a single SSS forms at the transition are likely very rare, in particular in disordered networks. Also, we do not expect the number of SSS that form at the transition to substantially affect our results.
	\item \emph{There are no $K$th-order zero modes at the transition, $N_{K\mathrm{th}}^\ast=0$, for $K>1$.} 
	We expect the occurrence of such modes to be rare. Moreover, some preliminary work indicates that allowing for $N_{K\mathrm{th}}^\ast>0$ for $K>1$ will mostly affect the prefactor for the entropic rigidity $\kappa_S$, but may not change our overall results (\aref{app:higher-order-delta-ell-Ns}).
	\item \emph{There are more than two first-order zero modes at the transition, $N_\mathrm{1st}^\ast>2$.}
	The marginal case of $N_\mathrm{1st}^\ast=1$ corresponds to the situation where there is no floppy-rigid transition (compare Fig.~\ref{fig:counting examples}), since no mechanisms would form in the floppy regime. It will be interesting to study the marginal case $N_\mathrm{1st}^\ast=2$ in future work.
	\item \emph{For our discussion in \aref{app:orders of magnitude delta r}, we needed the additional assumption in Eq.~\eqref{eq:deltar-rho}.} Intuitively, we believe that one can obtain our result in \aref{app:orders of magnitude delta r} quite generally even without this assumption, and it will be interesting to test if this is true in future research.
	\item \emph{There are no mechanisms at the transition.}
	The reason why we need this assumption is that with a degeneracy $\vecn{r}\,^\ast$ of the transition point configurations, in principle, the Taylor expansion in Eq.~\eqref{eq:Taylor-expansion} could depend on $\vecn{r}\,^\ast$.
	However, while properties like the singular values, $s_p$ with $p<N_s$, of the compatibility matrix $C_{in}^{(0)}$ may depend on $\vecn{r}\,^\ast$ \cite{Rocklin2018a,Mannattil2022}, our results should be unaffected as long as the SSS (more precisely $\tilde{w}_{N_s}$) and its behavior with respect to shear (more precisely $b_\varepsilon$) are independent of the choice of $\vecn{r}\,^\ast$.
	While this seems to be true intuitively, preliminary results suggest that stronger assumptions about the system than those listed in \autoref{eq:model under-constrained system} are required to rigorously show it. The tools required for such a proof are outside the scope of the ideas presented here, and their development is left for future work.
\end{itemize}
Thus, we expect that many of these criteria do not essentially restrict our results here and/or correspond to marginal cases. Yet, including these cases provides interesting avenues for future work.

In the future, it will be interesting to compare these analytical results to more numerical data \cite{Ninarello2022}.
Furthermore, as we have shown, all of the three parameters, $\kappa_E$, $\kappa_S$, and $b_\varepsilon$, are properties of the SSS that is created at the transition. Thus, better understanding the structure of this SSS will be important going forward.
Recently, conformal invariance has been harnessed to derive scaling properties of rigidity percolation \cite{Javerzat2023,Javerzat2024}. It will be interesting to see whether similar approaches can be used to understand the structure of the SSS in under-constrained materials, at least for some classes of disordered networks.
Finally, it will be interesting to understand how our work generalizes to systems that are constructed by adding to an under-constrained system a number of weak generalized springs to make it over-constrained \cite{Rens2018c,Lerner2023}.

\begin{acknowledgments}
We thank Chris Santangelo, Jen Schwarz, and Manu Mannattil for fruitful discussions.
We thank the Centre Interdisciplinaire de Nanoscience de Marseille (CINaM) for providing office space.
The project leading to this publication has received funding from France 2030, the French Government program managed by the French National Research Agency (ANR-16-CONV-0001), and from the Excellence Initiative of Aix-Marseille University - A*MIDEX.
\end{acknowledgments}

\appendix
\section{Implementation of periodic boundary conditions and shear}
\label{app:PBCs and shear}
As mentioned in \autoref{eq:model under-constrained system}, our framework is valid both for simple and pure shear. Here, we provide examples for the implementation of either, by providing explicit expressions for the spring length functions $L_i(\vecn{R},\mathcal{V},\gamma)$ for a system of ordinary linear springs.

To implement simple shear in a $D=2$-dimensional system, we compute the length of any spring $i$ that connects node $a_i$ to node $b_i$. We first define the components of the spring length vector $\vecd{L}_i=(L_{i,x},L_{i,y})$ as follows \cite{Note1}:
\begin{equation}
	\begin{aligned}
		L_{i,x}(\vecn{R},\mathcal{V},\gamma) &= R_{b_i,x} - R_{a_i,x} + (q_{i,x} + \gamma q_{i,y})\sqrt{\mathcal{V}} \\
		L_{i,y}(\vecn{R},\mathcal{V},\gamma) &= R_{b_i,y} - R_{a_i,y} + q_{i,y}\sqrt{\mathcal{V}},
	\end{aligned}\label{eq:simple shear}
\end{equation}
where $\vecd{R}_a=(R_{a,x},R_{a,y})$ is the position of some node $a$. The spring length is then given by: $L_i(\vecn{R},\mathcal{V},\gamma)=\vert \vecd{L}_i(\vecn{R},\mathcal{V},\gamma)\vert$.

In Eq.~\eqref{eq:simple shear}, the symbol $\vecd{q}_i=(q_{i,x},q_{i,y})\in\mathbb{Z}^D$ denotes the \emph{periodicity vector} of spring $i$ \cite{Merkel2014b,Merkel2018}.
A value of $q_{i,x}=0$ indicates that the spring does not cross the vertical periodic boundary of the system, while $q_{i,x}\neq0$ indicates that the spring \emph{does} cross the vertical periodic boundary.
In particular, a positive value $q_{i,x}>0$ indicates that when following the spring going from $\vecd{R}_{a_i}$ to $\vecd{R}_{b_i}$ one leaves via the right vertical boundary while entering via the left boundary exactly $q_{i,x}$ times.
Analogously, a negative value $q_{i,x}<0$ indicates that following the spring from $\vecd{R}_{a_i}$ to $\vecd{R}_{b_i}$ one leaves the via \emph{left} boundary while entering via the \emph{right} boundary exactly $-q_{i,x}$ times.

A similar way of implementing simple shear is given by:
\begin{equation}
	\begin{aligned}
		L_{i,x}(\vecn{R},\mathcal{V},\gamma) &= R_{b_i,x} - R_{a_i,x} + q_{i,x}\sqrt{\mathcal{V}} \\
		&\qquad + \gamma\Big(R_{b_i,y} - R_{a_i,y} + q_{i,y}\sqrt{\mathcal{V}}\Big)\\
		L_{i,y}(\vecn{R},\mathcal{V},\gamma) &= R_{b_i,y} - R_{a_i,y} + q_{i,y}\sqrt{\mathcal{V}}.
	\end{aligned}\label{eq:simple shear - Lees}
\end{equation}
This corresponds to Lees-Edward boundary conditions \cite{Lees1972}.

To implement pure shear in a $D=3$-dimensional system, we define the components of the length vector $\vecd{L}_i=(L_{i,x},L_{i,y},L_{i,z})$ of a spring $i$ that that connects node $a_i$ to node $b_i$ as follows:
\begin{equation}
	\begin{aligned}
		L_{i,x}(\vecn{R},\mathcal{V},\gamma) &= R_{b_i,x} - R_{a_i,x} + q_{i,x}e^\gamma\sqrt[3]{\mathcal{V}} \\
		L_{i,y}(\vecn{R},\mathcal{V},\gamma) &= R_{b_i,y} - R_{a_i,y} + q_{i,y}e^{-\gamma}\sqrt[3]{\mathcal{V}} \\
		L_{i,z}(\vecn{R},\mathcal{V},\gamma) &= R_{b_i,z} - R_{a_i,z} + q_{i,z}\sqrt[3]{\mathcal{V}}.
	\end{aligned}\label{eq:pure shear}
\end{equation}
Again, the spring length is then given by: $L_i(\vecn{R},\mathcal{V},\gamma)=\vert \vecd{L}_i(\vecn{R},\mathcal{V},\gamma)\vert$.
Different pure shear implementations can vary the $e^\gamma$ factors in the periodicity terms, such that the product of all these terms remains one. 

Note that in all example cases discussed here, the functions $L_i(\vecn{R},\mathcal{V},\gamma)$ are analytic as long as $L_i>0$ and $\mathcal{V}>0$.
Moreover, the relation Eq.~\eqref{eq:L-homogeneous} holds for all spring vector definitions, Eqs.~\eqref{eq:simple shear}--\eqref{eq:pure shear} with $d_i=1$.
Finally, a similar approach as in Eqs.~\eqref{eq:simple shear}--\eqref{eq:pure shear} can be used to implement simple or pure shear in vertex models instead of simple spring networks \cite{Merkel2014b,Merkel2018}.

\section{Orders of magnitude of the \texorpdfstring{$\Lambda_p$ and $\Delta\tilde{r}_q$}{dofs}}
\label{app:orders of magnitude delta r}
\subsection{Athermal limit}
\label{app:orders of magnitude delta r - athermal}
We discuss the scaling of the $\Delta\tilde{r}_q$ and the $\Lambda_p$ with $\varepsilon$ and $\gamma$ in the athermal limit at the energy minimum.  While in the floppy regime, where $E=0$, we have trivially $\Lambda_p=0$ for all $p=1,\dots,N_s$, we discuss here the rigid regime, where $E>0$.  For simplicity, we ignore the existence of zero eigenvalues of the matrix $\tilde{M}_{N_sqr}$ for $q,r=N_s,\dots,N_\mathrm{dof}$ (see also \autoref{sec:stiff-spring - zero modes}).

We focus here for clarity on the case where $\tilde{B}_p^{(1)}=\tilde{C}^{(1)}_{pq}=0$ for all $p,q$, for which Eq.~\eqref{eq:Lambdap} becomes:
\begin{equation}
	\Lambda_p
		= s_\underdot{p}\Delta\tilde{r}_\underdot{p} + \frac{1}{2}\tilde{M}_{pqr}\Delta\tilde{r}_q\Delta\tilde{r}_r
		 +\tilde{w}_p\varepsilon + \frac{1}{2}\tilde{B}_p^{(2)}\gamma^2.
	\label{eq:Lambdap-noB1}
\end{equation}
The general case of arbitrary $\tilde{B}_p^{(1)}$ and $\tilde{C}^{(1)}_{pq}$ is discussed in \aref{app:first-order-gamma}.

We focus on the limit of small $\varepsilon$ and $\gamma$.  
To simplify the subsequent calculations, we choose to approach this limit by Taylor expanding using the parameter $\eta\ll 1$ and setting $\varepsilon=\varepsilon_0\eta$ and $\gamma=\gamma_0\eta^{1/2}$. This way, for any given ratio $\varepsilon/\gamma^2=\varepsilon_0/\gamma_0^2$, we can reach the small $\varepsilon$ and $\gamma$ limit by decreasing $\eta$ accordingly.

We now discuss the scaling of the $\Delta\tilde{r}_q$ and the $\Lambda_p$ with $\eta$. 
We assume that $\Delta\tilde{r}_q$ can be expanded in power series of $\eta^{1/2}$, and it follows that the same is true for the $\Lambda_p$.
With coefficients $\rho_q^{(k/2)}$ and $\lambda_p^{(k/2)}$, we respectively have:
\begin{align}
	\Delta\tilde{r}_q &= \sum_{k=0}^\infty{\rho_q^{(k/2)}\eta^{k/2}}, \label{eq:deltar-rho}\\
	\Lambda_p &= \sum_{k=0}^\infty{\lambda_p^{(k/2)}\eta^{k/2}}.\label{eq:Lambda-lambda}
\end{align}
First, we find that $\rho_q^{(0)}=0$ for all $q=1,\dots,N_\mathrm{dof}$, because at the transition, where $\eta\sim\varepsilon+b_\varepsilon\gamma^2\rightarrow0$, all $\Delta\tilde{r}_q=0$. 
Then, insertion of Eqs.~\eqref{eq:deltar-rho} and \eqref{eq:Lambda-lambda} into Eq.~\eqref{eq:Lambdap-noB1} yields:
\begin{align}
	\lambda_p^{(0)} &= 0 \\
	\lambda_p^{(1/2)} &= s_\underdot{p}\rho_\underdot{p}^{(1/2)} \label{eq:lambda1/2}\\
	\lambda_p^{(1)} &= s_\underdot{p}\rho_\underdot{p}^{(1)} + \frac{1}{2}\tilde{M}_{pqr}\rho_q^{(1/2)}\rho_r^{(1/2)} + \mathrm{const.} \label{eq:lambda1}
\end{align}
To further discuss the values of $\rho_q^{(k/2)}$ and $\lambda_p^{(k/2)}$, we use the energy minimum condition:
\begin{equation}
	0 = \frac{1}{E_0}\frac{\partial E}{\partial \Delta\tilde{r}_q}\qquad\text{for $q=1,\dots,N_\mathrm{dof}$.}
\end{equation}
From this we obtain using Eqs.~\eqref{eq:energy-dimensionless-Lambda} and \eqref{eq:Lambdap-noB1}:
\begin{align}
	0 &=s_\underdot{q}\Lambda_\underdot{q}
	+\Lambda_p\tilde{M}_{pqr}\Delta\tilde{r}_{r} && \text{for $q=1,\dots,N_s-1$.}
	\label{eq:energy-minimum-p}\\
	0
	&= \Lambda_p\tilde{M}_{pqr}\Delta\tilde{r}_{r} && \text{for $q=N_s,\dots,N_\mathrm{dof}$.}
	\label{eq:energy-minimum-n}
\end{align}
On the right-hand side of both equations, sums over $p=1,\dots,N_s$ and $r=1,\dots,N_\mathrm{dof}$ are implied.

To order $\eta^{1/2}$, Eq.~\eqref{eq:energy-minimum-p} implies that $\lambda_p^{(1/2)}=0$ for $p=1,\dots,N_s-1$. With Eq.~\eqref{eq:lambda1/2}, this implies $\rho_p^{(1/2)}=0$ for $p=1,\dots,N_s-1$. Furthermore, since $s_{N_s}=0$, Eq.~\eqref{eq:lambda1/2} also implies that $\lambda_{N_s}^{(1/2)}=0$.

To order $\eta^1$, Eq.~\eqref{eq:energy-minimum-p} also implies that $\lambda_p^{(1)}=0$ for $p=1,\dots,N_s-1$. With Eq.~\eqref{eq:lambda1} follows that
\begin{equation}
	0 = s_\underdot{p}\rho_\underdot{p}^{(1)} + \frac{1}{2}\tilde{M}_{pqr}\rho_q^{(1/2)}\rho_r^{(1/2)} + \mathrm{const.} \quad~~ \text{for $p<N_s$.} \label{eq:lambdap=0}
\end{equation}
Since $\rho_q^{(1/2)}=0$ for $q<N_s$, the implied sums in the second term effectively run only over $q,r=N_s,\dots,N_\mathrm{dof}$.

However, $\lambda_{N_s}^{(1)}$ is generally \emph{non-zero} with
\begin{equation}
	\lambda_{N_s}^{(1)} = \frac{1}{2}\tilde{M}_{N_sqr}\rho_q^{(1/2)}\rho_r^{(1/2)} + \mathrm{const.} 
\end{equation}
Again, since $\rho_q^{(1/2)}=0$ for $q<N_s$, the implied sums in the second term effectively run only over $q,r=N_s,\dots,N_\mathrm{dof}$.

Consequentially, up to quadratic order in $\eta$, the system energy is dominated by $\Lambda_{N_s}$, and Eq.~\eqref{eq:energy-dimensionless-Lambda} becomes $E = E_0\Lambda_{N_s}^2/2$. 

Finally, to order $\eta^{3/2}$, Eq.~\eqref{eq:energy-minimum-n} together with the fact that in the rigid regime $\lambda_{N_s}^{(1)}\neq0$ implies that $\rho_q^{(1/2)}=0$ for $q\geq N_s$, where we ignore here zero eigen vectors of the matrix $\tilde{M}_{N_sqr}$. Since we showed above that $\rho_q^{(1/2)}=0$ also for $q<N_s$, we conclude that $\rho_q^{(1/2)}=0$ for all $q=1,\dots,N_\mathrm{dof}.$

\subsection{Stiff-spring limit}
\label{app:orders of magnitude delta r - stiff-spring}
In the stiff-spring limit, the arguments are similar to the athermal case, just that here, we already know from Eq.~\eqref{eq:Omega-eigenbasis} that $\Lambda_p=0$ for all $p=1,\dots,N_s$.
With Eq.~\eqref{eq:lambda1/2} follows that $\rho_p^{(1/2)}=0$ for $p=1,\dots,N_s-1$.
As a difference to the athermal limit, where also the $\rho_q^{(1/2)}$ with $q=N_s,\dots,N_\mathrm{dof}$ vanish, they take a finite value in this case.
Hence, $\Delta\tilde{r}_p\sim\eta^1$ for $p<N_s$, and $\Delta\tilde{r}_q\sim\eta^{1/2}$ for $q\geq N_s$.
As a consequence, in all $\Lambda_p$ with $p=1,\dots,N_s$ all terms $\sim\Delta\tilde{r}_q\Delta\tilde{r}_r$ with $q<N_s$ or $r<N_s$ can be neglected.

\subsection{General case}
\label{app:orders of magnitude delta r - general}
In the general case, we start from the partition sum, Eq.~\eqref{eq:Z-dimensionless}, which reads with Eq.~\eqref{eq:energy-dimensionless-Lambda}:
\begin{equation}
	Z \sim \int{\left(\prod_{q=1}^{N_\mathrm{dof}}{\d\Delta\tilde{r}_q}\right) \exp{\left[-\frac{\beta E_0}{2}\sum_{p=1}^{N_s}{\Lambda_p^2}\right]}}.
\end{equation}
Using Eqs.~\eqref{eq:Lambda-lambda} and \eqref{eq:lambda1/2}, this transforms into:
\begin{equation}
	\begin{split}
		Z \sim 	&\int{\left(\prod_{q=1}^{N_\mathrm{dof}}{\d\Delta\tilde{r}_q}\right)} \\
		&\quad\times\exp{\left[-\frac{\beta E_0\eta}{2}\sum_{p=1}^{N_s}{s_p^2[\rho_p^{(1/2)}]^2}\right]} \\
		&\quad\times\exp{\left[-\beta E_0\eta^{3/2}\sum_{p=1}^{N_s}{s_p\rho_p^{(1/2)}\lambda_p^{(1)}}\right]} \\
		&\quad\times\exp{\left[-\frac{\beta E_0\eta^2}{2}\sum_{p=1}^{N_s}{\Big([\lambda_p^{(1)}]^2+2s_p\rho_p^{(1/2)}\lambda_p^{(3/2)}\Big)}\right]}\\
		&\quad\times\exp{\left[-\frac{\beta E_0}{2}\text{(higher-order terms in $\eta$)}\right]}
		.
	\end{split}
\end{equation}
For $k_BT\lesssim E_0[\varepsilon + b_\varepsilon\gamma^2]^2$, we have $\beta E_0\eta^1\gtrsim \eta^{-1}\gg1$, which leads to an exponential suppression of those terms in the integral where $\rho_p^{(1/2)}\neq 0$ for any $p=1,\dots,N_s-1$.
Consequentially, to lowest order, $\Delta\tilde{r}_p=\rho_p^{(1)}\eta$ for $p=1,\dots,N_s-1$, and $\Delta\tilde{r}_q=\rho_q^{(1/2)}\eta^{1/2}$ for $q=N_s,\dots,N_\mathrm{dof}$.
We thus obtain, using Eq.~\eqref{eq:lambda1}:
\begin{equation}
	\begin{split}
		Z &\sim \eta^{(N_\mathrm{dof}+N_s-1)/2}\\
		&\times\int{\left(\prod_{p=1}^{N_s-1}{\d\rho_p^{(1)}}\right)\left(\prod_{q=N_s}^{N_\mathrm{dof}}{\d\rho_q^{(1/2)}}\right)} \\
		&\quad\times\exp{\Bigg[-\frac{\beta E_0\eta^2}{2}\sum_{p=1}^{N_s}{\bigg(s_p\rho_p^{(1)}}} \\
				&\qquad\qquad~+ \frac{1}{2}\sum_{q,r=N_s}^{N_\mathrm{dof}}{\tilde{M}_{pqr}\rho_q^{(1/2)}\rho_r^{(1/2)}} + \mathrm{const.}\bigg)^2\Bigg].
	\end{split}
\end{equation}
Since the terms $\sim\eta^2$, are generally non-zero, in particular for $p=N_s$, we neglect in the exponential terms of higher order than  $\eta^2$.
Since in the exponential, any term with sum index $p$ does not depend on any of the $\rho_q^{(1)}$ with $q\neq p$ (and $q<N_s$), we can rearrange the integral into:
\begin{equation}
	\begin{split}
		Z &\sim \eta^{(N_\mathrm{dof}+N_s-1)/2}\\
		&\times\int{\left(\prod_{q=N_s}^{N_\mathrm{dof}}{\d\rho_q^{(1/2)}}\right)} \exp{\Bigg[-\frac{\beta E_0\eta^2}{2}}\\
			&\qquad\quad\times\bigg( \frac{1}{2}\sum_{q,r=N_s}^{N_\mathrm{dof}}{\tilde{M}_{N_sqr}\rho_q^{(1/2)}\rho_r^{(1/2)}} + \mathrm{const.}\bigg)^2\Bigg]\\
		&\qquad\times\prod_{p=1}^{N_s-1}{\Bigg(}\int{\d\rho_p^{(1)}} \exp{\Bigg[}-\frac{\beta E_0\eta^2}{2}\bigg(s_p\rho_p^{(1)} \\
			&\qquad\qquad +\frac{1}{2}\sum_{q,r=N_s}^{N_\mathrm{dof}}{\tilde{M}_{pqr}\rho_q^{(1/2)}\rho_r^{(1/2)}} + \mathrm{const.}\bigg)^2\Bigg]\Bigg).
	\end{split}
\end{equation}
This corresponds to Eq.~\eqref{eq:Z-decomposed} in the main text.

\section{Non-zero first-order terms in \texorpdfstring{$\gamma$: $\tilde{B}^{(1)}_p$ and $\tilde{C}^{(1)}_{pq}$}{shear strain}}
\label{app:first-order-gamma}
Again to simplify our discussion here, we ignore the existence of mechanisms, and more generally the possibility of zero modes of the matrix $\tilde{M}_{N_sqr}$ for $q,r=N_s,\dots,N_\mathrm{dof}$, i.e.\ this matrix is strictly positive definite.

To discuss the general case of non-vanishing $\tilde{B}^{(1)}_p$, we first simplify by applying shifts to both $\varepsilon$ and $\gamma$,
\begin{align}
	\bar\varepsilon &:= \varepsilon - \Delta\varepsilon \\
	\bar\gamma &:= \gamma - \Delta\gamma.
\end{align}
We define $\Delta\varepsilon$ and $\Delta\gamma$ such that in the athermal limit $(\bar\varepsilon,\bar\gamma)=(0,0)$ is on the transition line, and such that after carrying out the Taylor expansion in Eq.~\eqref{eq:Taylor-expansion-ebOfC} with respect to $\bar\gamma$ instead of $\gamma$, the linear-order $\bar\gamma$ term in $\Lambda_{N_s}$ vanishes at the energy minimum:
\begin{equation}
	\Lambda_{N_s} = \frac{1}{2}\tilde{M}_{N_sqr}\Delta\tilde{r}_q\Delta\tilde{r}_r + \tilde{w}_{N_s}\bar\varepsilon + \tilde{C}_{N_sq}^{(1)}\Delta\tilde{r}_q\bar\gamma + \frac{1}{2}\tilde{B}_{N_s}^{(2)}\bar\gamma^2.\label{eq:app:GNs}
\end{equation}
Note that the $\Lambda_p$ with $p<N_s$ do in general still have a linear-order term in $\bar\gamma$:
\begin{equation}
	\begin{split}
		\Lambda_p &= s_\underdot{p}\Delta\tilde{r}_\underdot{p} + \frac{1}{2}\tilde{M}_{pqr}\Delta\tilde{r}_q\Delta\tilde{r}_r \\
		&\qquad\quad  + \tilde{w}_{p}\bar\varepsilon + \tilde{B}_p^{(1)}\bar\gamma + \tilde{C}_{pq}^{(1)}\Delta\tilde{r}_q\bar\gamma +\frac{1}{2}\tilde{B}_p^{(2)}\bar\gamma^2.	
	\end{split}
	\label{eq:app:Gp}
\end{equation}
Also, the $\Delta\tilde{r}_q$ may also be redefined due to the shifts in $\varepsilon$ and $\gamma$ due to a different non-dimensionalisation according to \autoref{sec:dimensionless quantities} and a change in the compatibility matrix.
If $\Delta\varepsilon$ and $\Delta\gamma$ are sufficiently small, it can be shown that they are given by $\Delta\varepsilon=(\tilde{B}^{(1)}_{N_s})^2/4\tilde{w}_{N_s}^2\!b_\varepsilon$ and $\Delta\gamma=-\tilde{B}^{(1)}_{N_s}/2\tilde{w}_{N_s}\!b_\varepsilon$, where $\tilde{B}^{(1)}_{N_s}$ is the linear-order term before the shift, and $b_\varepsilon$ is given by Eq.~\eqref{eq:be-with-linear-order-terms} below.
The symmetry of the system with respect to the sign of $\gamma$ combined with the central limit theorem suggest that, in general, these linear terms scale with system size as $\tilde{B}^{(1)}_{N_s}\sim N_s^{-1/2}$. Indeed, we numerically found earlier for phanomized triangular networks that $\tilde{B}^{(1)}_{N_s}\sim\Delta\gamma\sim N_\mathrm{dof}^{-1/2}\sim N_s^{-1/2}$ \cite{Lee2022}.

We first discuss the athermal limit using arguments similar to \aref{app:orders of magnitude delta r - athermal}. Using $\bar\eta$ as small parameter and fixing $\bar\varepsilon\sim\bar\eta$ and $\bar\gamma\sim\bar\eta^{1/2}$, we use Eq.~\eqref{eq:energy-minimum-p} to order $\bar\eta^{1/2}$ to conclude that for any $p<N_s$ up to order $\bar\eta^{1/2}$:
\begin{equation}
	0 = \Lambda_p = s_\underdot{p}\Delta\tilde{r}_\underdot{p} + \tilde{B}^{(1)}_p\bar\gamma.
\end{equation}	
Thus, to order $\bar\eta^{1/2}$:
\begin{equation}
	\Delta\tilde{r}_p = -\frac{\tilde{B}^{(1)}_p}{s_p}\bar\gamma\qquad\text{for $p<N_s$.}
\end{equation}
Inserting this into Eq.~\eqref{eq:app:GNs} together with the diagonalization of the matrix $\tilde{M}_{N_sqr}$ with $q,r\geq N_s$, Eq.~\eqref{eq:diagonalization-M}, we obtain up to order $\bar\eta^1$:
\begin{equation}
	\Lambda_{N_s} = \frac{1}{2}\sum_{k=1}^{N_\mathrm{dof}-N_s+1}{\mu_k\Big(\Delta q_k - \Delta q^0_k\Big)^2} + \tilde{w}_{N_s}\big[\bar\varepsilon + b_\varepsilon\bar\gamma^2\big],\label{eq:GNs-with-delta-qk0}
\end{equation}
where we introduced
\begin{equation}
  \Delta q^0_k := -\frac{\bar\gamma}{\mu_k}\sum_{r\geq N_s}{v^k_r\zeta_r} \label{eq:delta-qk0}
\end{equation}
with
\begin{equation}
	\zeta_r := \tilde{C}_{N_sr}^{(1)} -\sum_{p<N_s}{\tilde{M}_{N_srp}\frac{\tilde{B}^{(1)}_p}{s_p}}, \label{eq:zeta_r}
\end{equation}
and
\begin{equation}
	\begin{split}
		b_\varepsilon &:= \frac{1}{2\tilde{w}_{N_s}}\vast[\tilde{B}^{(2)}_{N_s} + 
		\sum_{p,q<N_s}{\frac{\tilde{B}^{(1)}_p}{s_p}\tilde{M}_{N_spq}\frac{\tilde{B}^{(1)}_q}{s_q}} \\
		&\qquad
		- 2\sum_{p<N_s}{\tilde{C}_{N_sp}^{(1)}\frac{\tilde{B}^{(1)}_p}{s_p}}
		- \sum_{\substack{q,r\geq N_s\\k}}{\zeta_q\frac{v^k_qv^k_r}{\mu_k}\zeta_r} 
		\vast].
	\end{split}\label{eq:be-with-linear-order-terms}
\end{equation}
Remember that we assumed that the matrix $\tilde{M}_{N_sqr}$ for $q,r=N_s,\dots,N_\mathrm{dof}$ is strictly positive definite, i.e.\ $\mu_k>0$ for all $k$.
In practice, to numerically compute $b_\varepsilon$ we found that any modes $k$ with $\mu_k=0$ can just be excluded from the last sum, where we numerically found that for these $k$, also $\sum_{q\geq N_s}\zeta_qv_q^k=0$.

For the stiff-spring limit and the general case of finite spring stiffness and temperature, these arguments are essentially the same as for the athermal limit.
In the stiff-spring limit, for non-vanishing $\tilde{B}^{(1)}_p$ and $\tilde{C}^{(1)}_{pq}$, the hyper-ellipsoid that describes the possible system configurations is now centered at $(\Delta q_k^0)$ (compare Eqs.~\eqref{eq:ellNs-diagonal} and \eqref{eq:GNs-with-delta-qk0}). According to Eq.~\eqref{eq:delta-qk0} this ellipsoid center displaces linearly with shear strain $\bar\gamma$.

In the general case, a similar line of argument holds as in \aref{app:orders of magnitude delta r - general}. However, the quantities that become zero are not $\rho_p^{(1/2)}$ for $p<N_s$, but instead $\rho_p^{(1/2)}+(\tilde{B}^{(1)}_p/s_p)\bar\gamma/\bar\eta$.

Using the scaling arguments from \aref{app:orders of magnitude delta r}, one can see that any further terms in the Taylor expansion, Eq.~\eqref{eq:Taylor-expansion}, will not change our results.
Specifically, when allowing for finite $\tilde{B}^{(1)}_p$, we have that $\Delta\tilde{r}_q\sim\eta^{1/2}$ for all $q$, while $\Lambda_{N_s}\sim\eta^1$, and the $\Lambda_p$ with $p<N_s$ are of higher order than $\sim\eta^1$. Consequentially, any higher-order term in Eq.~\eqref{eq:Taylor-expansion} will only yield terms with order of at least $\sim\eta^{3/2}$.

\section{Relation between the Hessian and the matrix \texorpdfstring{$\tilde{M}_{N_sqr}$}{M}}
\label{app:Hessian-M}
We evaluate the Hessian of the system,
\begin{equation}
	\begin{split}
		H_{qr} 
		&:= \frac{\partial^2E}{\partial\Delta\tilde{r}_q \partial\Delta\tilde{r}_r},
	\end{split}
\end{equation}
\emph{in the rigid athermal regime}, at an energy minimum close to the transition point, where we assume for simplicity that $\tilde{B}_p^{(1)}=\tilde{C}_{pq}^{(1)}=0$ for all $p,q$.

Using Eq.~\eqref{eq:energy-dimensionless-Lambda}:
\begin{equation}
	\begin{split}
	H_{qr} 
	& = E_0\sum_{p=1}^{N_s}{\left(\frac{\partial \Lambda_p}{\partial\Delta\tilde{r}_q}\,\frac{\partial \Lambda_p}{\partial\Delta\tilde{r}_r} + \Lambda_p\frac{\partial^2 \Lambda_p}{\partial\Delta\tilde{r}_q\partial\Delta\tilde{r}_r}\right)}.
	\end{split}
\end{equation}
Further, according to our discussion in \aref{app:orders of magnitude delta r - athermal}, at an energy minimum, all $\Delta\tilde{r}_q$ are at least of order $\sim\eta^1$. Also, $\Lambda_p=0$ up to order $\sim\eta^1$ for $p<N_s$. Thus, we have up to order $\sim\eta^1$:
\begin{equation}
	\begin{split}
		H_{qr} 
		& = E_0\sum_{p=1}^{N_s-1}{s_p^2\delta_{pq}\delta_{pr}} \\
		&\quad + E_0\sum_{p=1}^{N_s-1}\sum_{s=1}^{N_\mathrm{dof}}{s_p\Delta\tilde{r}_s\Big(\delta_{pq}\tilde{M}_{psr} + \tilde{M}_{psq}\delta_{pr}\Big)} \\
		&\quad + E_0\Lambda_{N_s}\tilde{M}_{N_sqr}
		.\label{eq:H-M}
	\end{split}
\end{equation}
In this expression, the first term, which is of absolute order in $\varepsilon$ and $\gamma$, corresponds to the Hessian at the transition point. The second and third terms are of order $\sim\eta^1$.

For $q,r\geq N_s$ only the third term in Eq.~\eqref{eq:H-M} is nonzero. Hence, to lowest order in $\eta$, the $N_\mathrm{1st}^\ast$ nonzero eigen modes of $\tilde{M}_{N_sqr}$ for $q,r\geq N_s$, are those eigen modes of $H_{qr}$ that have zero eigen value at the transition point, and whose eigen values become nonzero in the rigid regime, scaling as $\sim\Lambda_{N_s}\sim\eta^1$.

\section{Partition sum for the stiff-spring limit}
\label{app:partition sum stiff-spring limit}
Care needs to be taken when computing the partition sum for systems with fixed constraints \cite{vanKampen1984}.
In our generalized spring networks, in the stiff-spring limit, all springs are imposed to be at their respective rest lengths. Yet, it is not immediately obvious how to impose these constraints to obtain the correct phase space volume. Naively, one may be tempted to explicitly impose the length of each spring, as we do in Eq.~\eqref{eq:Omega0}, to obtain a phase space volume:
\begin{equation}
	\Omega_\mathrm{def1} = \int{\left(\prod_{n=1}^{N_\mathrm{dof}}{\d R_n}\right)\prod_{i=1}^{N_\mathrm{s}}{\delta(L_i-L_{0i})}}. \label{eq:Omega_def1}
\end{equation}
Yet, another possibility would be to just impose that the energy should vanish:
\begin{equation}
	\Omega_\mathrm{def2} = \int{\left(\prod_{n=1}^{N_\mathrm{dof}}{\d R_n}\right)\delta(E)}.\label{eq:Omega_def2}
\end{equation}
After all, if the energy vanishes, all springs have to be at their respective rest length (compare Eq.~\eqref{eq:energy}).
However, both definitions, Eq.~\eqref{eq:Omega_def1} and Eq.~\eqref{eq:Omega_def2}, will lead to different results. This is because the integration of a Dirac delta involves the evaluation of the Jacobian of the argument of the Dirac delta with respect to the integration variable. Since the Jacobian of the energy creates additional prefactors, $K_i(L_i-L_{0i})$, the resulting phase space volumes $\Omega$ will differ between both definitions by a non-constant factor.

To obtain the correct expression for the phase space volume, we evaluate the partition sum
\begin{equation}
	Z = \int{\left(\prod_{n=1}^{N_\mathrm{dof}}{\d R_n}\right) \exp{\left(-\frac{\beta }{2}\sum_{i=1}^{N_s}{K_i\big(L_i-L_{0i}\big)^2}\right)} }
\end{equation}
in the stiff-spring limit.  A Gaussian converges to a Dirac Delta in the limit of small standard deviation:
\begin{equation}
	\lim_{K_i\rightarrow\infty}\sqrt{\frac{\beta K_i}{2\pi}}\exp{\left(-\frac{\beta K_i(L_i-L_{0i})^2}{2}\right)} = \delta(L_i-L_{0i}).
\end{equation}
Thus, whenever all $K_i$ are large enough such that for all $i=1,\dots,N_s$ the $(\beta K_i)^{-1/2}$ become much smaller than all relevant length scales of the system, the partition sum can be expressed as:
\begin{equation}
	\begin{split}
		Z 
		&= \left(\frac{2\pi}{\beta}\right)^{N_s/2}\prod_{i=1}^{N_s}{K_i^{-1/2}} \\
		&\qquad\qquad\times \int{\left(\prod_{n=1}^{N_\mathrm{dof}}{\d R_n}\right) \prod_{i=1}^{N_s}{\delta\big(L_i-L_{0i}\big)}}.
	\end{split}
\end{equation}
This indicates that the correct approach is to use Dirac deltas on the spring lengths rather than the energy $E$. Note that in Eq.~\eqref{eq:Omega0}, we have ignored the first prefactor that only depends on temperature, and so is not relevant for the elastic system properties.  Moreover, while the second prefactor does not depend on strain or temperature, it converges to zero for $K_i\rightarrow\infty$. However, to discuss the elastic system properties, we only need derivatives of $\log Z$, where this term only creates a constant albeit diverging offset.

\section{Higher-order dependencies of \texorpdfstring{$\Delta\tilde\ell_{N_s}$}{the SSS} on the dofs}
\label{app:higher-order-delta-ell-Ns}
Here we outline how some of our arguments change in the presence of $K$th-order zero modes with $K>1$.
In particular, we discuss how the phase space volume scaling with strain changes.

We start from the following specific generalization of the expansion of $\Delta\tilde\ell_{N_s}$ from Eqs.~\eqref{eq:Taylor-expansion-ebOfC}, \eqref{eq:ellNs-diagonal}:
\begin{equation}
	\Delta\tilde{\ell}_{N_s} = \sum_{K=1}^\infty\frac{1}{(K+1)!}\sum_{k=1}^{N_{K\mathrm{th}}^\ast}{\mu_{K,k}\Delta q_{K,k}^{K+1}} + \frac{1}{2}\tilde{B}_{N_s}^{(2)}\gamma^2. \label{eq:appE-deltalNs}
\end{equation}
Here, $(K+1)!$ denotes the faculty of $K+1$, the variables $\Delta q_{K,k}$ denote the amplitudes of the $K$th-order zero modes, and the associated coefficients are $\mu_{K,k}>0$. For simplicity, we set again $\tilde{B}_p^{(1)}=\tilde{C}^{(1)}_{pq}=0$, and we ignore mixed terms between the different $\Delta q_{K,k}$.

Assuming that the SSS mode $\Lambda_{N_s}$ still dominates the energy in this case, we find that $N_{K\mathrm{th}}^\ast=0$ for even $K$.  This is because otherwise, a $K$th-order zero mode for even $K$ would allow for states with $\Lambda_{N_s}=0$ for $\varepsilon>0$ and $\gamma=0$, even though the system is supposed to be rigid with $E>0$ in this regime.  $\Lambda_{N_s}$ can become zero for $N_{K\mathrm{th}}^\ast>0$ for even $K$, because a $K$th-order zero mode would introduce a term $\sim\Delta q_{K,k}^{K+1}$ with an odd exponent, and consequentially, $\Delta q_{K,k}$ can be adjusted by the energy minimization such that $\Lambda_{N_s}=0$.  Taken together, this implies that the sum over $K$ in Eq.~\eqref{eq:appE-deltalNs} actually only includes the odd integers.

Thus, the phase space volume is governed by an integral of the following kind:
\begin{equation}
	\begin{split}
		&\Omega_{N_s} \sim I(\varepsilon,\gamma)
		:=
		\int{\prod_{K=1}^\infty\prod_{k=1}^{N_{K\mathrm{th}}^\ast}{\d \Delta q_{K,k}}} \\
		&\times\delta\left(\sum_{K=1}^\infty\frac{1}{(K+1)!}\sum_{k=1}^{N_{K\mathrm{th}}^\ast}{\mu_{K,k}\Delta q_{K,k}^{K+1}} + \tilde{w}_{N_s}\Big[\varepsilon + b_\varepsilon\gamma^2\Big]\right),
		\label{eq:Omega-remaining-integrals-app}
	\end{split}
\end{equation}
To extract the dependency of $I$ on $\varepsilon$ and $\gamma$, we first define:
\begin{equation}
	\begin{split}
	I_1
		&:=
		\int{\prod_{K=1}^\infty\prod_{k=1}^{N_{K\mathrm{th}}^\ast}{\d x_{K,k}}} \\
		&\qquad\quad\times\delta\left(\sum_{K=1}^\infty\frac{1}{(K+1)!}\sum_{k=1}^{N_{K\mathrm{th}}^\ast}{\mu_{K,k}x_{K,k}^{K+1}} + \tilde{w}_{N_s}\right),
	\end{split}
\end{equation}
\emph{which depends neither on $\varepsilon$ nor on $\gamma$}.
Now we can transform the expression in Eq.~\eqref{eq:Omega-remaining-integrals-app} using the following substitutions for all integration variables:
\begin{equation}
	\Delta q_{K,k} = \big[\varepsilon + b_\varepsilon\gamma^2\big]^{1/(K+1)}x_{K,k},\label{eq:N-Omega-substitution}
\end{equation}
which yields:
\begin{equation}
	\Omega_{N_s}(\varepsilon, \gamma)\sim I(\varepsilon, \gamma) = \big[\varepsilon + b_\varepsilon\gamma^2\big]^{N_\Omega}I_1
\end{equation}
with:
\begin{equation}
	N_\Omega = -1 + \sum_{K=1}^\infty{\frac{N_{K\mathrm{th}}^\ast}{K+1}}.\label{eq:NOmega}
\end{equation}
The first term in $N_\Omega$ arises from removing the factor $[\varepsilon + b_\varepsilon\gamma^2]$ from the Dirac delta, and all the remaining terms arise from the substitution of the integration variables.
In this argument, we have excluded modes that leave $\Delta\tilde\ell_{N_s}$ invariant. They do not contribute to $N_\Omega$.  This is because any substitution like in Eq.~\eqref{eq:N-Omega-substitution}, while creating a prefactor, also rescales the integral bounds by the same prefactor, both of which exactly cancel each other out. Meanwhile, for any mode that \emph{does} affect $\Delta\tilde\ell_{N_s}$, the rescaling of the integral bounds does not matter, since the integration domain of these modes is effectively limited by the Dirac delta, not the integral bounds.

Note that for the case discussed in the main text, i.e.\ for $N_\mathrm{1st}^\ast>0$ and $N_{K\mathrm{th}}^\ast=0$ for $K>1$, we have indeed $N_\Omega=(N_\mathrm{1st}^\ast-2)/2$, consistent with Eq.~\eqref{eq:NOmega}. Taken together, as indicated by Eq.~\eqref{eq:NOmega}, also higher-order zero modes at the transition may contribute to the exponent $N_\Omega$; they are just attenuated according to their order $K$.
If the SSS mode still dominates the phase space volume scaling, we also have $\kappa_S=N_\Omega k_B/D\mathcal{V}^\ast$.

\section{Tension and shear modulus in the general case}
\label{app:derivatives general case}
To obtain the isotropic tension $t:=(\partial F/\partial \varepsilon)/D\mathcal{V}$, we first use the chain rule. To lowest order, $\mathcal{V}=\mathcal{V}^\ast$, and so we obtain from Eq.~\eqref{eq:general-F}:
\begin{equation}
	\begin{split}
		t = \frac{1}{D\mathcal{V}^\ast}\Bigg[&\frac{\partial \bar{F}}{\partial\hat\varepsilon_S}\frac{\partial\hat\varepsilon_S}{\partial\varepsilon} + \frac{\partial \bar{F}}{\partial\varepsilon} \\
		&\qquad+\frac{1}{2\beta\bar{F}''}\left(\frac{\partial \bar{F}''}{\partial\hat\varepsilon_S}\frac{\partial\hat\varepsilon_S}{\partial\varepsilon} + \frac{\partial \bar{F}''}{\partial\varepsilon}\right)\Bigg].
	\end{split}
\end{equation}
We compute to lowest order:
\begin{align}
	\frac{\partial\hat\varepsilon_S}{\partial\varepsilon}
		&= -\frac{\hat\varepsilon_S + b_\varepsilon\gamma^2}{\left\vert\varepsilon + b_\varepsilon\gamma^2\right\vert\sqrt{1+\theta}}\\
	\frac{1}{D\mathcal{V}^\ast}\frac{\partial\bar{F}}{\partial\hat\varepsilon_S}
		&=  -\kappa_E(\varepsilon-\hat\varepsilon_S) - \frac{\kappa_ST}{\hat\varepsilon_S+b_\varepsilon\gamma^2} = 0 \label{eq:derivative-barF-hateS}\\
	\frac{1}{D\mathcal{V}^\ast}\frac{\partial\bar{F}}{\partial\varepsilon}
		&= \kappa_E(\varepsilon-\hat\varepsilon_S) \\
	\frac{1}{D\mathcal{V}^\ast}\frac{\partial\bar{F}''}{\partial\hat\varepsilon_S}
		&=  -\frac{2\kappa_ST}{(\hat\varepsilon_S+b_\varepsilon\gamma^2)^3}\\
	\frac{1}{D\mathcal{V}^\ast}\frac{\partial\bar{F}''}{\partial\varepsilon}
		&= 0.
\end{align}
We thus obtain:
\begin{equation}
	\begin{split}
		t
		&= \kappa_E(\varepsilon-\hat\varepsilon_S)\left[1 + \frac{1}{2(N_\mathrm{1st}^\ast-2)}\,\frac{\theta}{1+\theta}\right], \label{eq:general-t1-app}
	\end{split}
\end{equation}
where we have also used that
\begin{equation}
	\bar{F}'' = -D\mathcal{V}^\ast\kappa_E\frac{\left\vert\varepsilon + b_\varepsilon\gamma^2\right\vert\sqrt{1+\theta}}{\hat\varepsilon_S + b_\varepsilon\gamma^2}
\end{equation}
and
\begin{equation}
	\kappa_E(\varepsilon-\hat\varepsilon_S) = - \frac{\kappa_ST}{\hat\varepsilon_S+b_\varepsilon\gamma^2}, \label{eq:kEkS-e}
\end{equation}
which follows from Eq.\eqref{eq:hat-eS}. The term $\sim(N_\mathrm{1st}^\ast-2)^{-1}$ in Eq.~\eqref{eq:general-t1-app} corresponds to the terms involving $\bar{F}''$.

For the shear modulus $G:=(\partial^2F/\partial\gamma^2)/\mathcal{V}$, we have to lowest order:
\begin{equation}
	\begin{split}
		G
		= \frac{1}{\mathcal{V}^\ast}\Bigg[&\frac{\partial^2\bar{F}(\hat\varepsilon_S(\varepsilon, \gamma); \varepsilon, \gamma)}{\partial\gamma^2} \\
		&\qquad+ \frac{1}{2\beta}\frac{\partial^2}{\partial\gamma^2}\log{\bar{F}''(\hat\varepsilon_S(\varepsilon, \gamma); \varepsilon, \gamma)}\Bigg]. \label{eq:G-appendix-derivative}
	\end{split}
\end{equation}
We compute with Eqs.~\eqref{eq:derivative-barF-hateS} and \eqref{eq:kEkS-e}:
\begin{equation}
	\begin{split}
		\frac{1}{\mathcal{V}^\ast}\frac{\partial\bar{F}(\hat\varepsilon_S(\varepsilon, \gamma); \varepsilon, \gamma)}{\partial\gamma} 
		&= \frac{1}{\mathcal{V}^\ast}\frac{\partial\bar{F}(\hat\varepsilon_S; \varepsilon, \gamma)}{\partial\gamma} \\
		&= -\frac{2D\kappa_STb_\varepsilon\gamma}{\hat\varepsilon_S+b_\varepsilon\gamma^2} \\
		&= 2D\kappa_E(\varepsilon-\hat\varepsilon_S)b_\varepsilon\gamma,
	\end{split}
\end{equation}
and further:
\begin{equation}
	\begin{split}
		&	\frac{1}{\mathcal{V}^\ast} \frac{\partial^2\bar{F}(\hat\varepsilon_S(\varepsilon, \gamma); \varepsilon, \gamma)}{\partial\gamma^2} \\
		&= 	\frac{1}{\mathcal{V}^\ast}\Bigg[\frac{\partial^2\bar{F}(\hat\varepsilon_S; \varepsilon, \gamma)}{\partial\gamma\partial\hat\varepsilon_S}\frac{\partial\hat\varepsilon_S}{\partial\gamma} + \frac{\partial^2\bar{F}(\hat\varepsilon_S; \varepsilon, \gamma)}{\partial\gamma^2}\Bigg] \\
	\end{split}
\end{equation}
with
\begin{align}
	\frac{1}{\mathcal{V}^\ast}\frac{\partial^2\bar{F}(\hat\varepsilon_S; \varepsilon, \gamma)}{\partial\gamma\partial\hat\varepsilon_S} &= -2D\kappa_Eb_\varepsilon\gamma \\
	\frac{\partial\hat\varepsilon_S}{\partial\gamma} &=
	-b_\varepsilon\gamma\left(1+\frac{\varepsilon + b_\varepsilon\gamma^2}{\vert\varepsilon + b_\varepsilon\gamma^2\vert\sqrt{1+\theta}}\right) \\
	&= -2b_\varepsilon\gamma\frac{\varepsilon - \hat\varepsilon_S}{\vert\varepsilon + b_\varepsilon\gamma^2\vert\sqrt{1+\theta}} \\
	\frac{1}{\mathcal{V}^\ast}\frac{\partial^2\bar{F}(\hat\varepsilon_S; \varepsilon, \gamma)}{\partial\gamma^2} &= 2D\kappa_E(\varepsilon-\hat\varepsilon_S)b_\varepsilon.
\end{align}
Thus:
\begin{equation}
	\begin{split}
		&\frac{1}{\mathcal{V}^\ast}\frac{\partial^2\bar{F}(\hat\varepsilon_S(\varepsilon, \gamma); \varepsilon, \gamma)}{\partial\gamma^2} \\
		&= 2D\kappa_E(\varepsilon-\hat\varepsilon_S)b_\varepsilon \Bigg(1 + \frac{2b_\varepsilon\gamma^2}{\vert\varepsilon + b_\varepsilon\gamma^2\vert\sqrt{1+\theta}}\Bigg).
	\end{split}\label{eq:G-barF-g2}
\end{equation}
Furthermore:
\begin{align}
		&\frac{\partial\bar{F}''(\hat\varepsilon_S(\varepsilon, \gamma); \varepsilon, \gamma)}{\partial\gamma}  \nonumber\\
		&= 	\frac{\partial\bar{F}''(\hat\varepsilon_S; \varepsilon, \gamma)}{\partial\hat\varepsilon_S}\frac{\partial\hat\varepsilon_S}{\partial\gamma} + \frac{\partial\bar{F}''(\hat\varepsilon_S; \varepsilon, \gamma)}{\partial\gamma} \nonumber\\
		&= D\mathcal{V}^\ast\kappa_E\Bigg[\left(\frac{1}{\hat\varepsilon_S +b_\varepsilon\gamma^2} - \frac{\hat\varepsilon_S-\varepsilon}{(\hat\varepsilon_S +b_\varepsilon\gamma^2)^2}\right) \nonumber\\
		&\qquad\qquad\qquad\quad\times\left(-2b_\varepsilon\gamma\frac{\varepsilon - \hat\varepsilon_S}{\vert\varepsilon + b_\varepsilon\gamma^2\vert\sqrt{1+\theta}}\right) \nonumber\\
		&\qquad\qquad\quad -2b_\varepsilon\gamma\frac{\hat\varepsilon_S-\varepsilon}{(\hat\varepsilon_S +b_\varepsilon\gamma^2)^2}\Bigg] \nonumber\\
		&= -2D\mathcal{V}^\ast\kappa_Eb_\varepsilon\gamma\frac{\varepsilon-\hat\varepsilon_S}{(\hat\varepsilon_S +b_\varepsilon\gamma^2)^2\vert\varepsilon + b_\varepsilon\gamma^2\vert\sqrt{1+\theta}} \nonumber\\
		&\qquad\times\Big[\varepsilon + b_\varepsilon\gamma^2 - \vert\varepsilon + b_\varepsilon\gamma^2\vert\sqrt{1+\theta}\Big] \nonumber\\
		&= -4D\mathcal{V}^\ast\kappa_Eb_\varepsilon\gamma\frac{\varepsilon-\hat\varepsilon_S}{(\hat\varepsilon_S +b_\varepsilon\gamma^2)\vert\varepsilon + b_\varepsilon\gamma^2\vert\sqrt{1+\theta}},
\end{align}
and thus:
\begin{equation}
	\begin{split}
		\frac{\partial}{\partial\gamma} \log{\bar{F}''(\hat\varepsilon_S(\varepsilon, \gamma); \varepsilon, \gamma)} = 4b_\varepsilon\gamma\frac{\varepsilon-\hat\varepsilon_S}{(\varepsilon + b_\varepsilon\gamma^2)^2(1+\theta)}.
	\end{split}
\end{equation}
The second derivative is:
\begin{equation}
	\begin{split}
		&\frac{\partial^2}{\partial\gamma^2} \log{\bar{F}''(\hat\varepsilon_S(\varepsilon, \gamma); \varepsilon, \gamma)} \\
		&= \frac{4b_\varepsilon(\varepsilon-\hat\varepsilon_S)}{(\varepsilon + b_\varepsilon\gamma^2)^2(1+\theta)} \\
		&\quad+ \frac{4b_\varepsilon\gamma}{(\varepsilon + b_\varepsilon\gamma^2)^2(1+\theta)}\times 2b_\varepsilon\gamma\frac{\varepsilon - \hat\varepsilon_S}{\vert\varepsilon + b_\varepsilon\gamma^2\vert\sqrt{1+\theta}} \\
		&\quad - \frac{16b_\varepsilon^2\gamma^2(\varepsilon-\hat\varepsilon_S)}{(\varepsilon + b_\varepsilon\gamma^2)^3(1+\theta)^2} \\
		&= \frac{4b_\varepsilon(\varepsilon-\hat\varepsilon_S)}{(\varepsilon + b_\varepsilon\gamma^2)^2(1+\theta)} \Bigg[1 + \frac{2b_\varepsilon\gamma^2}{\vert\varepsilon + b_\varepsilon\gamma^2\vert\sqrt{1+\theta}} \\
		&\qquad\qquad\qquad\qquad\qquad\quad\;\; - \frac{4b_\varepsilon\gamma^2}{(\varepsilon + b_\varepsilon\gamma^2)(1+\theta)}\Bigg].
	\end{split}\label{eq:G-barFpp-g2}
\end{equation}
From Eqs.~\eqref{eq:kappaS} and \eqref{eq:kEkS-e} follows:
\begin{equation}
	\begin{split}
		\frac{1}{2\beta \mathcal{V}^\ast} = \frac{D\kappa_ST}{N_\mathrm{1st}^\ast-2} = \frac{D\kappa_E}{N_\mathrm{1st}^\ast-2}\,\theta\frac{(\varepsilon + b_\varepsilon\gamma^2)^2}{4}
	\end{split}\label{eq:beta-prefactor}
\end{equation}
Combining Eqs.~\eqref{eq:G-appendix-derivative}, \eqref{eq:G-barF-g2}, \eqref{eq:G-barFpp-g2}, and \eqref{eq:beta-prefactor}, we obtain Eq.~\eqref{eq:G-general} in the main text.

\bibliography{references.bib}

\end{document}